\documentclass[%
 reprint,
superscriptaddress,
 amsmath,amssymb,longbibliography,
 aps,
 prl,
]{revtex4-1}
\pdfoutput=1
\usepackage{ulem}
\usepackage{hyperref}
\usepackage{graphicx}
\usepackage{amsmath}
\usepackage{color}
\usepackage{physics}
\usepackage{dsfont}
\usepackage{comment}
\usepackage{physics}
\usepackage{booktabs}
\usepackage{pifont}
\usepackage[T1]{fontenc}
\usepackage{makecell}
\usepackage{pdfpages} 
\usepackage{pgffor} 

\makeatletter
\AtBeginDocument{\let\LS@rot\@undefined}
\makeatother

\def\supplementfilename{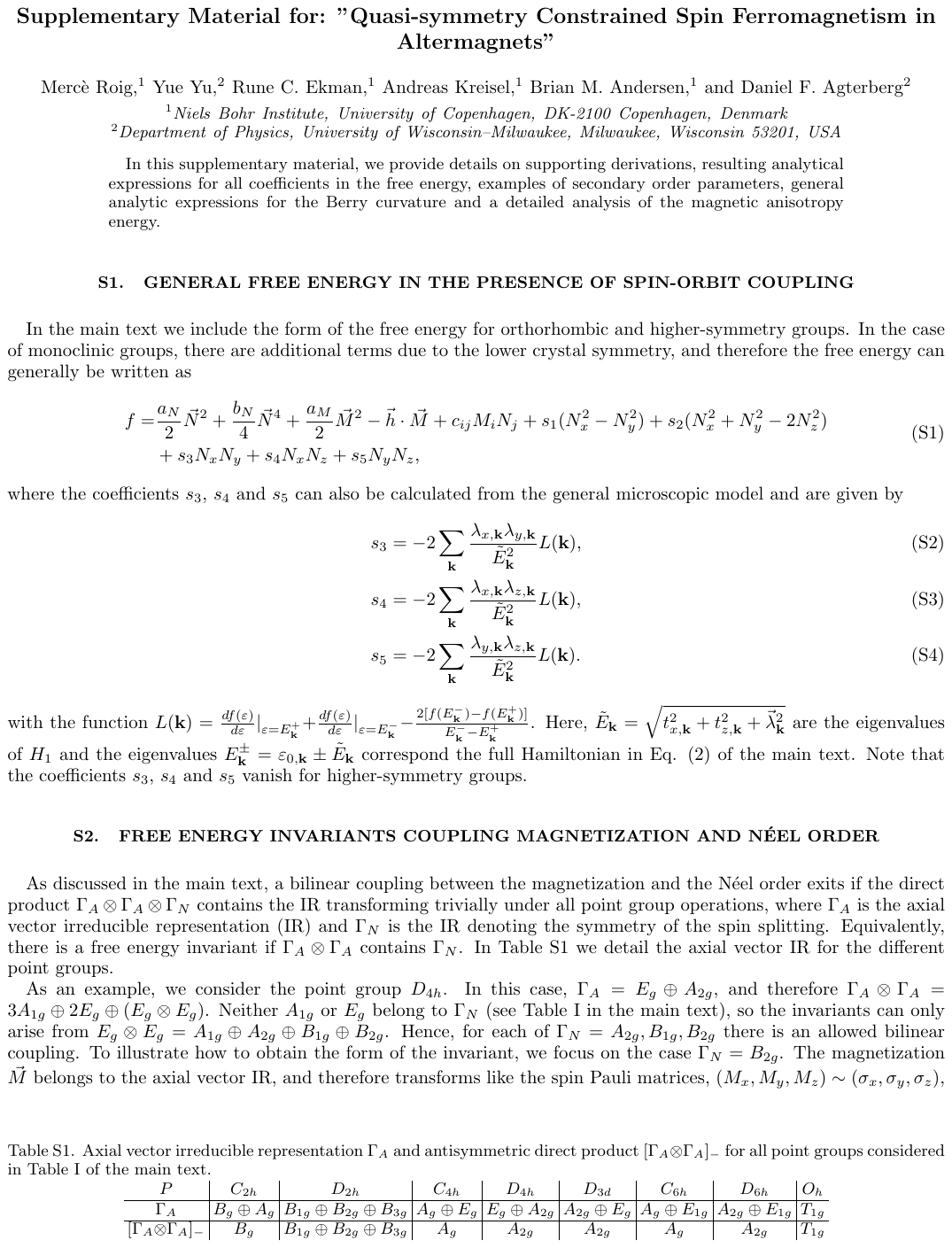}
\def\numbersupplementpages{\the\pdflastximagepages}

\newif\ifarXiv
\arXivtrue 

\definecolor{limegreen}{rgb}{0.2, 0.8, 0.2}
\definecolor{orange}{rgb}{1.0, 0.5, 0.0}

\definecolor{emerald}{rgb}{0.31, 0.78, 0.47}
\definecolor{blue(ncs)}{rgb}{0.0, 0.53, 0.74}

\def\red{\color{red}}

\hypersetup{
     colorlinks=true,
     linkcolor=magenta,
     filecolor=blue,
     citecolor=emerald,      
     urlcolor =blue(ncs),
}

\newcommand{\kv}{{\bf k}}

\newcommand{\Nvec}{{\vec{N}}}
\newcommand{\Mvec}{{\vec{M}}}

\begin{document}
\title{Quasi-symmetry Constrained Spin Ferromagnetism in Altermagnets}

\author{Mercè Roig}
\affiliation{Niels Bohr Institute, University of Copenhagen, DK-2100 Copenhagen, Denmark}
\affiliation{Department of Physics, University of Wisconsin–Milwaukee, Milwaukee, Wisconsin 53201, USA} 

\author{Yue Yu}
\affiliation{Department of Physics, University of Wisconsin–Milwaukee, Milwaukee, Wisconsin 53201, USA} 

\author{Rune C. Ekman}
\affiliation{Niels Bohr Institute, University of Copenhagen, DK-2100 Copenhagen, Denmark} 

\author{Andreas Kreisel}
\affiliation{Niels Bohr Institute, University of Copenhagen, DK-2100 Copenhagen, Denmark} 

\author{Brian M. Andersen}
\affiliation{Niels Bohr Institute, University of Copenhagen, DK-2100 Copenhagen, Denmark} 

\author{Daniel F. Agterberg}
\affiliation{Department of Physics, University of Wisconsin–Milwaukee, Milwaukee, Wisconsin 53201, USA}


\vskip 1cm

\begin{abstract}
Altermagnets break time-reversal symmetry and their spin-orbit coupling (SOC) allow for an anomalous Hall effect (AHE) that depends on the direction of the Néel ordering vector. The AHE and the ferromagnetic spin moment share the same symmetry and hence are usually proportional. However, density functional theory (DFT) calculations find that the AHE exists with negligible ferromagnetic spin moment for some compounds, whereas it reaches sizable values for other altermagnets. By examining realistic minimal models for altermagnetism in which the DFT phenomenology is captured, we uncover a general SOC-enabled quasi-symmetry, the uniaxial spin space-group, that provides a natural explanation for the amplitude of the ferromagnetic spin moment across the vast range of different altermagnetic materials. Additionally, we derive analytic expressions for the magnetic anisotropy energy, providing a simple means to identify the preferred Néel vector orientation for altermagnets.
\end{abstract}

\maketitle

\textit{{\red Introduction.--}} 
The Hall effect has been a frontier theme in condensed matter physics for more than half a century. Fundamental understanding of the quantized Hall conductance and the anomalous Hall effect (AHE) has helped pave the way for topological classifications of quantum matter and the importance of Berry curvature in transport properties~\cite{Nagaosa2010,Sodemann2015,Gao2023,Kaplan2024,Fang2024,Du2021,Wang2023}. The discovery of altermagnetism (AM) and its associated large antiferromagnetic AHE is yet another testimony to this development~\cite{Lee2024Jan,Krempasky2024Feb,Osumi2024Mar,Li2024May,Yang2024May,Reimers2024Mar,Ding2024,Feng2022Nov, Kluczyk2023Oct,GonzalezBetancourt2023Jan}, which may open new opportunities for devices utilizing dissipationless transport and spintronics technologies~\cite{Smejkal2020Jun,Smejkal2022Sep,Smejkal2022Dec,Smejkal2022Jun,Fang2024,Ma2021May}.

The AHE in AM, which we describe here as an intra-unit cell Néel order, is only nonzero for certain directions of the Néel vector and in the presence of relativistic spin-orbit coupling (SOC). This, in addition, induces weak ferromagnetism (FM), which satisfies the same symmetry properties as the AHE \cite{McClarty2023Aug,Xiao2024}. This implies that, in principle, FM and AHE share the same dependence on the orientation of the Néel vector~\cite{Smejkal2022Jun,Fernandes2024Jan,Cheong2024Jan}. However, from DFT it is found that the relative amplitude of the AHE transport coefficient and the FM spin moment depends strongly on the particular AM material under investigation. For example, for  RuO$_2$ and MnTe, DFT finds a large AHE of 40~S/cm and 300 S/cm, respectively, while RuO$_2$ has a nearly zero FM moment (that we estimate is of the order $10^{-7} \mu_B$) and Mn in MnTe has $10^{-4} \mu_B$~\cite{Smejkal2020Jun,Mazin2024Jul,GonzalezBetancourt2023Jan}. In the case of MnTe and RuO$_2$, weak FM was also reported experimentally~\cite{Berlijn2017Feb,Kluczyk2023Oct}, though the AM ground state of the latter material is still debated~\cite{Hiraishi2024Apr,Kessler2024May}. A weak FM spin moment has also been identified by DFT in CrSb~\cite{Autieri2023Dec}. By contrast, in other AM, such as FeSb$_2$, DFT finds a large AHE of 143 S/cm~\cite{Mazin2021Oct,Attias:2024}, and our calculations estimate a moderate FM moment of 0.03~$\mu_B$. Similarly, the material RuF$_4$ has also been predicted by DFT to exhibit a large FM spin moment of 0.22 $\mu_B$ on the Ru sites~\cite{Milivojevic2024May}. These results suggest that while the AHE is generically large when allowed by symmetry, the corresponding FM moment is strongly dependent on the AM state and this is poorly understood.

To gain deeper insight into the interplay between the AHE, the FM moment, and the AM state it is crucial to analyze the role of SOC using realistic models. Here, by examining  realistic microscopic models for AM in which the DFT phenomenology is captured~\cite{Roig2024Feb}, we identify which AM states have a large SOC induced AHE but a small FM spin moment and find good  agreement with DFT results. In addition, we find that our model-based results are more generally valid. Specifically, they apply to all microscopic theories in which AM appears as a consequence of the exchange interaction. To show this, we identify a general SOC-enabled quasi-symmetry \cite{Li2024,Guo2022}, a uniaxial spin-space group,  that  establishes for which AM symmetries the symmetry-allowed FM is large or vanishingly small. Finally, the presence of SOC in AM breaks the spin-space degeneracy and leads to a preferred direction for the Néel vector. For example, DFT calculations have revealed that the moments are orthogonal to the a-b plane in RuO$_2$~\cite{Feng2022Nov}, while in  MnTe, FeSb$_2$, Nb$_2$FeB$_2$ and Ta$_2$FeB$_2$ the moments are predicted to be in-plane~\cite{Mazin2021Oct,Mazin2024Jul,Hou2023}.
By examining analytic expressions for the Landau coefficients derived from realistic microscopic models, we elucidate how the structure of SOC allows for the AM anisotropy energy to be understood.

\textit{{\red Interplay between magnetization and AM Néel order.--}} 
Introducing $\Nvec$ as the intra-unit cell Néel order and $\Mvec$ as the magnetization, the general form of the free energy density in the presence of SOC can be written as
\begin{align}
    F = & \frac{a_{N}}{2} \Nvec^2 + \frac{b_{N}}{4} \Nvec^4 + \frac{a_{M}}{2} \Mvec^2 - \vec{h} \cdot \Mvec      \label{eq:free_energy_SOC}  \\
    & +  c_{ij} M_i N_j  + s_1 (N_x^2 - N_y^2) + s_2 (N_x^2 + N_y^2 - 2N_z^2), \nonumber
\end{align}
where $a_{N}$, $b_{N}$, $a_{M}$ and $c_{ij}$ are  temperature-dependent Landau coefficients, $\vec{h}$ is an applied field, and $s_1$ and $s_2$ are the coefficients determining the magnetic anisotropy energy due to SOC.
Note that the bilinear coupling between the two orders $M_i$ and $N_j$ is only allowed in the presence of SOC~\cite{McClarty2023Aug}. Without SOC, the spin-space group rotations, $[R_S||R_G]$, with spin space ($R_S$) and real space group operations ($R_G$) are uncoupled, which forces the bilinear coupling to vanish since $\vec{N}$ is non-trivial under pure $R_G$ operations. In the magnetic space group, the rotations act simultaneously on both spaces, and therefore $R_S$ and the rotation portion of $R_G$ must be the same, allowing this bilinear coupling to be non-zero. Eq.~(\ref{eq:free_energy_SOC}) describes orthorhombic or higher-symmetry point groups, for monoclinic groups see the supplementary material (SM)~\cite{Supplementary}.

\begin{table}[t]
\caption{Lowest order free energy invariant between $M_i$ and $N_j$ in the presence of SOC for different point groups $P$ and IRs of the spin splitting $\Gamma_N$, with the function $f_{\Gamma_N}(\kv)$ transforming as $\Gamma_N$. The last column indicates if the $M_i$  generated from the AM order parameter is linear order in the SOC, denoted by a check mark, or higher order in SOC, denoted by a cross.}
\label{tab:mJcoupling_SOC}
\begin{tabular}{c|c|c|c|c}
\hline \hline
$P$ &  $\Gamma_N$ & $f_{\Gamma_N}(\kv)$ & \makecell{Lowest order\\invariant} & \makecell{$M_i$ linear \\ in SOC}\\ \hline 
$C_{2h}$ & $B_g$ & $\alpha k_x k_z + \beta k_y k_z$ & \makecell{$\alpha_1 N_x M_z$, $\alpha_2 N_y M_z$ \\ $\alpha_3 N_zM_y$, $\alpha_4 N_z M_x$} & \ding{51} \\ \hline
$D_{2h}$ & $B_{1g}$ & $k_x k_y$ & $\alpha_1 M_x N_y + \alpha_2 M_y N_x$ &  $\textrm{\ding{51}}$ \\
$D_{2h}$ & $B_{2g}$ & $k_x k_z$ & $\alpha_1 M_y N_z + \alpha_2 M_z N_y$ & $\textrm{\ding{51}}$ \\
$D_{2h}$ & $B_{3g}$ & $k_y k_z$ & $\alpha_1 M_z N_x + \alpha_2 M_x N_z$ & $\textrm{\ding{51}}$\\ \hline
$C_{4h}$ & $B_{g}$ & \makecell{$\alpha(k_x^2 - k_y^2)$ \\ + $\beta k_x k_y$} & \makecell{$M_x N_y + M_y N_x$,  \\$M_xN_x - M_yN_y$} & \ding{56} \\ \hline
$D_{4h}$ & $A_{2g}$ & $k_x k_y (k_x^2-k_y^2)$ & $M_x N_y - M_y N_x$ & \ding{51} \\
$D_{4h}$ & $B_{1g}$ & $k_x^2 - k_y^2$ & $M_x N_x - M_y N_y$ & \ding{56} \\
$D_{4h}$ & $B_{2g}$ & $k_x k_y$ & $M_x N_y + M_y N_x$ & \ding{56}  \\ \hline
$D_{3d}$ & $A_{2g}$ & $k_x k_z (k_x^2 - 3 k_y^2)$ & $M_x N_y - M_y N_x$ & \ding{51} \\ \hline
$C_{6h}$ & $B_{g}$ & \makecell{$\alpha k_y k_z (k_y^2 - 3k_x^2)$\\ ${+}\beta k_x k_z(k_x^2-3k_y^2)$} & \makecell{$\alpha M_z N_y(3N_x^2-N_y^2)$ \\ ${+}\beta M_z N_x (3N_y^2-N_x^2)$} & \ding{56} \\ \hline
$D_{6h}$ & $A_{2g}$ & \makecell{$k_x k_y(k_x^2 - 3k_y^2)$\\$\times(k_y^2-3k_x^2)$} & $M_x N_y - M_y N_x$ & \ding{51} \\
$D_{6h}$ & $B_{1g}$ & $k_y k_z(3k_x^2 - k_y^2)$ & $M_zN_y(3N_x^2-N_y^2)$ & \ding{56} \\
$D_{6h}$ & $B_{2g}$ & $k_xk_z(k_x^2-3k_y^2)$ & $M_zN_x(N_x^2-3N_y^2)$ & \ding{56} \\ \hline
$O_{h}$ & $A_{2g}$ & \makecell{$k_x^4(k_y^2 -k_z^2)$ \\ $+k_y^4(k_z^2 - k_x^2)$ \\ $+k_z^4(k_x^2-k_y^2)$} & \makecell{$ M_xN_x (N_y^2-N_z^2)$\\ $+ M_y N_y (N_z^2-N_x^2)$ \\ $+ M_z N_z (N_x^2-N_y^2)$} & \ding{56} \\
\hline \hline
\end{tabular}
\end{table}

The lowest order invariant coupling $M_i$ and $N_j$ can be determined from symmetry analysis. The magnetization $\Mvec$ belongs to the axial vector irreducible representation (IR) $\Gamma_A$, see SM~\cite{Supplementary}. In contrast, $\Nvec$ belongs to the IR $\Gamma_A \otimes \Gamma_N$, with $\Gamma_N$ the IR denoting the symmetry of the AM spin splitting. Hence, a coupling of $M_i$ and $N_j$ exists if the direct product $\Gamma_A\otimes\Gamma_A \otimes \Gamma_N$ contains the IR transforming trivially under all point group operations or, equivalently, there is a free energy invariant if $\Gamma_A \otimes \Gamma_A$ contains $\Gamma_N$~\cite{McClarty2023Aug}.
In Table~\ref{tab:mJcoupling_SOC} we provide the form of the free energy invariants considering the relevant point groups and symmetries for the spin splitting identified in Ref.~\cite{Roig2024Feb}. Note that for $C_{6h}$, $O_h$ and certain IRs $\Gamma_N$ of $D_{6h}$, a bilinear coupling is not allowed, and therefore the coupling is of higher order. Following a similar procedure, the lowest-order coupling between the magnetization and the Néel vector can also be identified~\cite{Hecker2024Jun,Supplementary}, and is included in Table~\ref{tab:mJcoupling_SOC}.

\textit{{\red Microscopic models.--}} To investigate the dependence of the induced FM spin moment on the SOC strength, we initially consider the general form of the minimal model for AM from Ref.~\cite{Roig2024Feb}, and carry out a model-independent quasi-symmetry based analysis later. Thus, we start from the normal state Hamiltonian 
\begin{equation}
    H_0  = \varepsilon_{0,\kv} + t_{x,\kv}\tau_x+t_{z,\kv}\tau_z+\tau_y \Vec{\lambda}_{\kv} \cdot \vec{\sigma},
    \label{eq:minimal_model}
\end{equation}
where $\tau_i$ represent sublattice and $\sigma_i$ spin degrees of freedom. The specific form of the parameters entering the model depends on the space group, the point group and the Wyckoff site symmetry. Here, $t_{x,\kv}$ is an inter-sublattice hopping term, $\varepsilon_{0,\kv}$ is the sublattice independent dispersion, and $\vec{\lambda}_\kv$ is the SOC. The crystal asymmetric hopping term $t_{z,\kv}$ exhibits a $\kv$ dependence that transforms as the non-trivial IR $\Gamma_N$ since it describes the local symmetry breaking from multipole moments ~\cite{Hayami2018Oct,Bhowal2024Feb,Roig2024Feb}.

Initially, we analyze the AHE. For the Néel order along the an arbitrary direction $l$, $N_l$, the only contribution to the Berry curvature that is linear in SOC originates from the SOC component parallel to $l$, that is $\lambda_l$, and is given by~\cite{Roig2024Feb,Supplementary}
\begin{align}
    \Omega_{\alpha, \beta, ij}^{(N_l)} = & \frac{1}{2E_{\alpha,\beta}^3} \sum_{m,n = i,j} \varepsilon_{mn} \Big[(N_l+\beta |t_{z,\kv}|) \partial_m \lambda_{l,\kv} \partial_n t_{x,\kv}  \notag\\
    & + \text{sgn}(t_{z,\kv}) \beta\, t_{x,\kv} \partial_m t_{z,\kv} \partial_n \lambda_{l,\kv} \notag\\
    & +  \text{sgn}(t_{z,\kv}) \beta\, \lambda_{l,\kv} \partial_m t_{x,\kv} \partial_n t_{z,\kv}\Big],
    \label{eq:general_Berrycurv_analytic}
\end{align}
where the dispersion is $E_{\alpha=\pm,\beta=\pm} = \alpha\Big(N_l^2+ \lambda_{l,\kv}^2 + t_{x,\kv}^2 + t_{z,\kv}^2 + 2\beta|N_l
| |t_{z,\kv}|\Big)^{1/2}$.
This expression is generically non-zero.
Thus, when allowed by symmetry, the AHE is expected to be large and linear in SOC for all point groups and AM symmetries. In the SM~\cite{Supplementary}, we carry out calculations of the AHE that support this conclusion.

To study now the interplay of the Néel order $\Nvec$ and the induced $\Mvec$, we consider the perturbation
\begin{equation}
    H' = \tau_z \Nvec \cdot \Vec{\sigma} + \Mvec\cdot\vec{\sigma}
    \label{eq:perturbation_interaction}
\end{equation}
 to the normal state Hamiltonian.
We calculate the corrections to the normal state free energy close to the critical temperature as the magnetic order sets in by evaluating the loop expansion of the free energy. To second order it reads
\begin{align}
    F^{(2)} = \frac{1}{2\beta} \sum_{i\omega_n} \Tr [{G}_0 H' {G}_0 H'].
    \label{eq:f_quadratic}
\end{align}
Here, the bare Green's function projected to the band basis is given by
\begin{equation}
    {G}_0(\kv,i\omega_n) = \sum_{a=\pm} G_0^{a}(\kv,i\omega_n) \ket{u^{a}_\kv} \bra{u_\kv^{a}}, 
    \label{eq:G_projected_bandbasis}
\end{equation}
where $G_0^{(\pm)}(\kv,i\omega_n) =  \frac{1}{i\omega_n - (\varepsilon_{0,\kv}\pm \tilde E_{\kv})}$ denotes the Green's function in the band basis, with the two-fold degenerate eigenenergies ${E}_\kv^{\pm} = \varepsilon_{0,\kv} \pm \tilde E_{\kv}$ with $\tilde E_{\kv}=\sqrt{t_{x,\kv}^2 + t_{z,\kv}^2 + \Vec{\lambda}_{\kv}^2 }$. The projection operator $P_\kv^a = \ket{u^{a}_\kv} \bra{u_\kv^{a}}$ in Eq.~\eqref{eq:G_projected_bandbasis} transforms from the sublattice basis onto band $a$ at wavevector $\kv$~\cite{Graf2021Aug,Supplementary}.

To examine the bilinear coupling between $\Nvec$ and $\Mvec$, we derive an expression for the coefficient $c_{ij}$ in Eq.~\eqref{eq:free_energy_SOC}. The analytic expressions for the other coefficients in Eq.~\eqref{eq:free_energy_SOC} can be found in the SM~\cite{Supplementary}. At one-loop level, the quadratic free energy contribution in Eq.~\eqref{eq:f_quadratic} coupling $\Mvec$ and $\Nvec$ is given by
\begin{equation}
    F_{N\!M}^{(2)}\!\! = \!\! \frac{1}{\beta}\! \Tr \! \!\Bigg[ \!  \sum_{a,b,i\omega_n}  \! G^a(\kv,i\omega_n) G^b(\kv,i\omega_n)\tau_z \Nvec\cdot \Vec{\sigma}  P^a_{\kv} \Mvec\cdot\vec{\sigma} P^b_{\kv} \Bigg].
\end{equation}
Calculating the trace and performing the Matsubara frequency sum, we obtain
\begin{equation}
    \begin{aligned}
    F_{N\!M}^{(2)} =  &2 \sum_\kv  \frac{t_{x,\kv}}
    {\tilde E^2_\kv }L(\kv)
    \Vec{\lambda}_\kv  \cdot (\Mvec \cross \Nvec),
    \end{aligned}
    \label{eq:coef_eps_mJ}
\end{equation}
with the function 
\begin{align}
    L(\kv)= \frac{df(\varepsilon)}{d\varepsilon}\Bigg|_{\varepsilon = E_\kv^+} {+} \frac{df(\varepsilon)}{d\varepsilon}\Bigg|_{\varepsilon = E_\kv^-} {-} \frac{2[f(E_\kv^-) {-} f(E_\kv^+)]}{E_\kv^- {-} E_\kv^+},
    \label{eq:intravsinter}
\end{align}
incorporating the density of states and Lindhard term.
The combination $(\Mvec \cross \Nvec)$ in Eq.~\eqref{eq:coef_eps_mJ} reveals that a nonzero invariant exists only if the antisymmetric direct product of the two axial IRs $[\Gamma_A \otimes \Gamma_A]_{-}$ contains $\Gamma_N$. In Table~\ref{tab:mJcoupling_SOC} we list whether the invariant can be generated to linear order of SOC, the antisymmetric product for the different point groups is detailed in the SM~\cite{Supplementary}. The SOC is expected to be a weak effect in AM~\cite{Smejkal2022Sep}, and therefore the induced magnetization will be vanishingly small when it is not generated to linear order. As seen from Table~\ref{tab:mJcoupling_SOC}, for the point group $D_{4h}$ the SOC-linear invariant is only generated for $\Gamma_N = A_{2g}$.
Focusing on crystals with rutile structure i.e. space group (SG) 136 and Wyckoff position 2a for the magnetic atoms as in  RuO$_2$, MnF$_2$, NiF$_2$, and CoF$_2$, we have $\Gamma_N = B_{2g}$. Consequently, the induced $\Mvec$ is at least quadratic in SOC, as opposed to the material candidates Nb$_2$FeB$_2$ and Ta$_2$FeB$_2$, which have $\Gamma_N = A_{2g}$.
Notably, Table~\ref{tab:mJcoupling_SOC} also shows that the FM moment induced in orthorhombic materials ($D_{2h}$) is generally expected to be larger. 

\begin{figure}[bt]
\begin{center}
\includegraphics[angle=0,width=.95\linewidth]{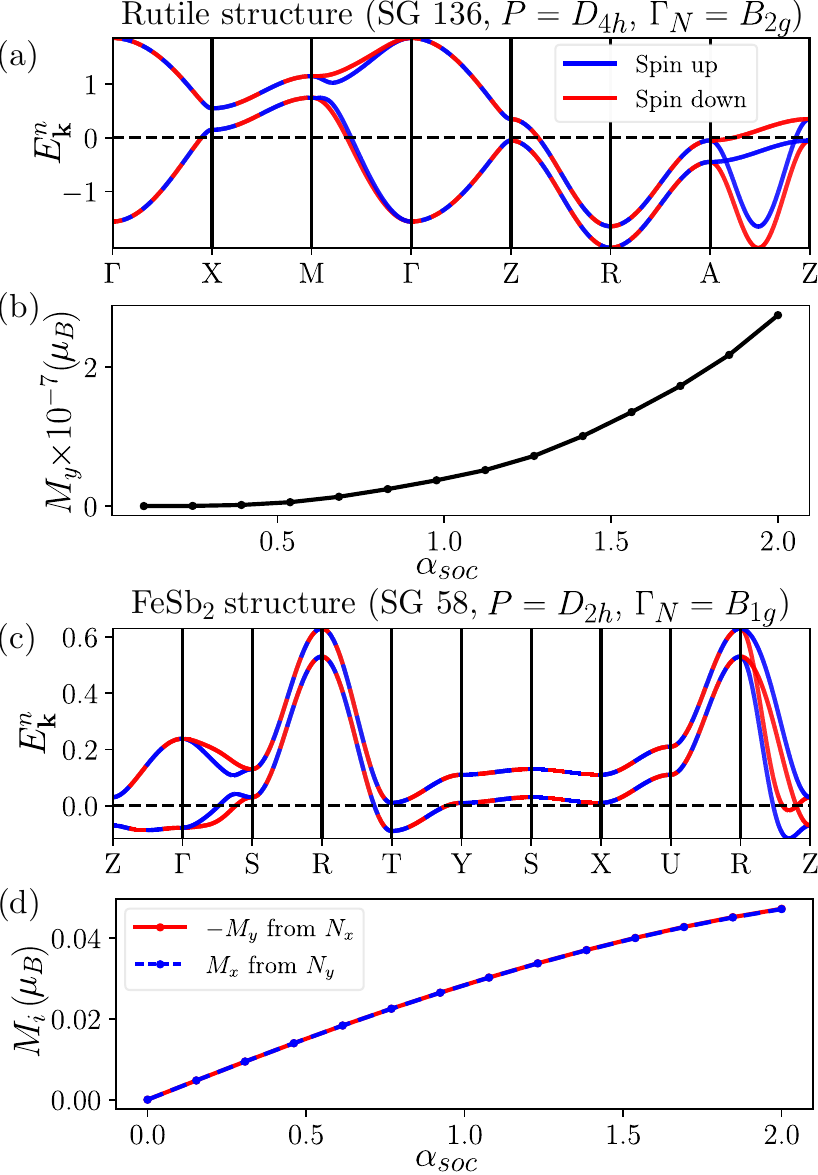}
\caption{Minimal model band structures and induced magnetization as a function of SOC strength for RuO$_2$ (a)-(b) vs FeSb$_2$ (c)-(d). We take $\vec{\lambda} = \alpha_{soc} \vec{\lambda}_0$, with $\vec\lambda_0=(0.05,0.05,0.17)\,\textrm{eV}$ and $N_x=0.2$ for RuO$_2$, and $\vec\lambda_0=(2.7,6.6,75)\,\textrm{meV}$ and $N_x = 0.05$ for FeSb$_2$, both estimated from DFT results (see SM~\cite{Supplementary}). 
In agreement with Table~\ref{tab:mJcoupling_SOC}, $\Mvec$  is induced along the $y$ axis for $N_x$, and scales quadratically (linearly) with the SOC strength for RuO$_2$ (FeSb$_2$). In (d) we show also the case with the Néel vector along the $y$-axis $N_y$, inducing $M_x \simeq -M_y$.}
\label{fig:inducedM}
\end{center}
\end{figure}

To further verify these points, in Fig.~\ref{fig:inducedM} we show the calculated
$\Mvec=\mu_B\sum_{a,\kv}\bra{u^{a}_\kv}\vec{S} \ket{u^{a}_\kv}f(E^a_{\kv})$
relevant for (a)~rutile structure ($D_{4h}$, $\Gamma_N = B_{2g}$) and (b)~FeSb$_2$ structure ($D_{2h}$, $\Gamma_N = B_{1g}$). As seen, the SOC-induced $\Mvec$ indeed scales quadratically (linearly) with the SOC strength for SG 136 (FeSb$_2$) and is significantly smaller for the band relevant for rutile AM  compared to FeSb$_2$. For the case of FeSb$_2$ we additionally compare the induced $\vec{M}$ for the Néel vector along the $x$-axis and the $y$-axis in Fig.~\ref{fig:inducedM}(d) to see if $\vec{M}$ follows the predicted microscopic $(M_xN_y-M_yN_x)$ result for the invariant. Indeed, even though $M_x$ and $M_y$ are not symmetry-related in this orthorhombic system, they are of opposite sign and nearly identical in magnitude for this material. In summary, these results explicitly demonstrate the properties summarized in Table \ref{tab:mJcoupling_SOC}, which can be applied to gain similar insight for many classes of AM materials. 

Finally, we note that in order to understand the quadratic dependence of $\vec{M}$ on SOC from exact diagonalization, when $\vec{M}$ is not allowed to linear order, requires the inclusion of secondary order parameters~\cite{McClarty2023Aug}, equivalent to two-loop calculations ~\cite{Supplementary}. Specifically, when AM order sets in, secondary order parameters are also induced by symmetry, and they can give rise to a finite coupling between the two orders $\Mvec$ and $\Nvec$. Such secondary order parameters include other spin textures as well as a current loop order, see SM~\cite{Supplementary}. The free energy for a secondary order parameter $\vec{O}$ can be written as
\begin{equation}
    F = \gamma^{(1)} \vec{O}^2 + \gamma^{(2)}_{ij} N_i {O}_j + \gamma^{(3)}_{ij} M_i O_j,
\end{equation}
which couples bilinearly to $\Mvec$ and $\Nvec$. Thus, $\Nvec$ can also induce a magnetization $\Mvec$ through the secondary order parameter $\vec{O}$, and our minimal models reveal that in this case the FM spin moment is at least quadratic in SOC, as discussed in the SM~\cite{Supplementary}. 

\begin{figure}[t]
\begin{center}
\includegraphics[angle=0,width=.95\linewidth]{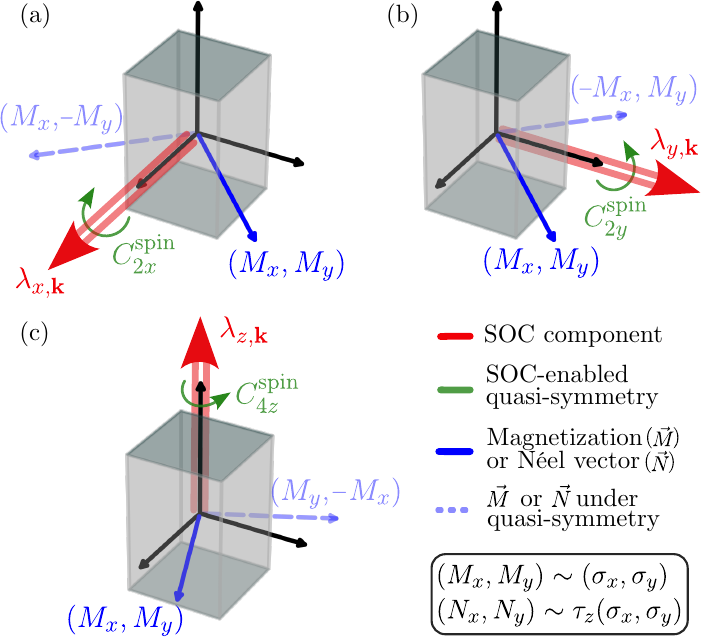}
\caption{SOC-enabled quasi-symmetry in a tetragonal system when only a single SOC component (a) $\lambda_{x,\kv}$, (b) $\lambda_{y,\kv}$ or (c) $\lambda_{z,\kv}$ is present. The quasi-symmetry acts only in spin space, transforming the magnetization $\Mvec$ and the spin components of the Néel vector $\Nvec$ in the same way.}
\label{fig:quasisym}
\end{center}
\end{figure}

\textit{{\red Quasi-symmetry protection of negligible FM.--}} Our minimal models are in agreement with the DFT results, which suggests a more general explanation beyond specific loop expansions or microscopic models. Indeed, it is possible to understand the above results using the recently introduced concept of quasi-symmetry~\cite{Li2024,Guo2022}, describing emergent approximate symmetries when certain terms in the Hamiltonian become negligible. Here we use SOC to generate our quasi-symmetry. Specifically, we consider the quasi-symmetry that emerges when two of the SOC components ($\lambda_x,\lambda_y,$ or $\lambda_z$) vanish, as illustrated in Fig.~\ref{fig:quasisym} for a tetragonal system. The resulting symmetry  group, which we denote the uniaxial spin-space group due to its spin rotational invariance around the SOC direction, has higher symmetry than the magnetic space group but lower symmetry than the spin space group. This is relevant for determining which Landau coefficients are linear in one of $\lambda_{x,y,z}$. For example, for $\lambda_z\ne 0$ with $\lambda_x=\lambda_y=0$, the normal state Hamiltonian gains additional symmetries, i.e., quasi-symmetries. Importantly, whether the $\lambda_z$-linear contribution to the Landau coefficient is permitted depends not only on the intrinsic symmetries of the crystal, but also on these emergent quasi-symmetries~\cite{Li2024,Guo2022}.

Here, we apply this to the Landau coefficients of $M_xN_y$ and $M_yN_x$. When only $\lambda_x$ ($\lambda_y$) SOC is present, the normal state Hamiltonian acquires an additional two-fold spin-rotational symmetry $[C_{2x}||E]$ ($[C_{2y}||E]$), see Fig.~\ref{fig:quasisym}(a) (Fig.~\ref{fig:quasisym}(b)), which prohibits $\lambda_x$ (or $\lambda_y$)-linear contribution to these two Landau coefficients since $M_x$ ($N_x$) is odd under this quasi-symmetry while $N_y$ ($M_y$) is even. When only $\lambda_z$ is present, a relevant quasi-symmetry in the uniaxial spin point group is the four-fold spin-rotational symmetry $[C_{4z}||E]$. As seen from Fig.~\ref{fig:quasisym}(c), under this symmetry the SOC-linear contribution to $M_xN_y$ and $M_yN_x$ coefficients must have opposite sign. This behavior is universal and applies to any space group, regardless of whether it has intrinsic 4-fold rotational symmetry, as demonstrated in Fig.~\ref{fig:inducedM}(d) for FeSb$_2$. In structures with SG 136, however, its intrinsic tetragonal crystal symmetries force these two coefficients to be identical at all orders of SOC. Consequently, the SOC-linear contribution must vanish, as shown in Fig.~\ref{fig:inducedM}(b). The complete discussion on other space groups can be found in SM~\cite{Supplementary}. This also implies that for MnTe and CrSb, which have point group $D_{6h}$ and $\Gamma_N = B_{1g}$, the FM moment will be cubic in SOC (see End Matter), independent of the non-linear coupling terms in Table~\ref{tab:mJcoupling_SOC}, in agreement with Refs.~\cite{Mazin2024Jul,Autieri2023Dec}. Hence, the SOC enabled quasi-symmetry demonstrates that the results of the microscopic model are very general, and naturally explains when the FM spin moment is large or small depending on the AM symmetry. The only assumption on the microscopic Hamiltonian that underlies this analysis is that AM is an instability purely in the spin-channel, that is, it is driven only by exchange interactions without any orbital angular momentum contribution.

The relationship between the SOC direction and non-vanishing AHE in Eq.~\eqref{eq:general_Berrycurv_analytic} can also be understood using SOC-based quasi-symmetry arguments like those given above, which do not depend on the form of the microscopic model. Specifically, the AHE measures current and voltage response, and hence is even under spin-rotational symmetries. For the N\'eel order along $\hat{x}$, when only the $\lambda_{y}$ ($\lambda_z$) SOC is kept, the N\'eel order is odd under the resulting quasi-spin-rotational symmetry $[C_{2y}||E]$ ($[C_{2z}||E]$). This prevents a $\lambda_y$ ($\lambda_z$)-linear SOC contribution to the AHE. When  only the $\lambda_x$ SOC is kept, the N\'eel order is even under the resulting $[C_{2x}||E]$ symmetry, and therefore the $\lambda_x$-linear contribution to AHE is allowed by the quasi-symmetry. It naturally follows that the AHE is given by the SOC component that is parallel to the N\'eel vector.

\textit{{\red Magnetic anisotropy energy.--}}
The presence of SOC leads to a preferred direction for the Néel vector. As seen from the free energy in Eq.~\eqref{eq:free_energy_SOC}, this is captured by the  $s_1$ and $s_2$ Landau coefficients. Thus, deriving  analytic expressions for these coefficients from the microscopic models is useful to provide insight into the easy axis direction. Using the general microscopic model in Eq.~\eqref{eq:minimal_model} and focusing on the quadratic free energy correction due to SOC (see Eq.~\eqref{eq:f_quadratic}), the coefficients can be written as
\begin{align}
    &s_1 = - \frac{1}{2} \sum_\kv \frac{\lambda_{x,\kv}^2{-}\lambda_{y,\kv}^2}{\tilde E^2_\kv
    } L(\kv)
    , \label{eq:MAE_s1} \\
    &s_2 =  - \frac{1}{6} \sum_\kv \frac{ \lambda_{x,\kv}^2{+}\lambda_{y,\kv}^2{-}2\lambda_{z,\kv}^2}{\tilde E^2_\kv
    }L(\kv),
    \label{eq:MAE_s2}
\end{align}
with the function $L(\kv)$ defined in Eq.~\eqref{eq:intravsinter}. 
The sign of these coefficients fixes the easy axis for a specific AM material. For instance, focusing on SG 136, $\lambda_{z,\kv}$ can be ignored as it is  smaller than $\lambda_{x,\kv}$ and $\lambda_{y,\kv}$~\cite{Roig2024Feb}. Hence, in general we expect that $s_2>0$ and, as a consequence, the easy axis is out of plane.
However, Eqs.~\eqref{eq:MAE_s1}-\eqref{eq:MAE_s2} also reveal that the moment direction may switch depending on the Fermi energy. In particular, when the Lindhard function (interband term in $L(\kv)$) dominates over the density of states (intraband term), $L(\kv)$ can change sign leading to $s_2<0$ and in-plane moment orientation. The switch of the AM moments as a function of the Fermi energy from the $c$-axis to the in-plane direction has been reported in Ref.~\cite{Feng2022Nov} for RuO$_2$. As shown in the SM, calculations of $s_1$ and $s_2$ for a minimal model of RuO$_2$ indeed yield $s_1=0$, $s_2>0$. In addition, application to FeSb$_2$ reveal that $s_1>0$, $s_2<0$, i.e. moments aligned along the in-plane $y$-axis, in agreement with DFT studies~\cite{Mazin2021Oct}. A systematic application of the magnetic anisotropy energy based on  Eqs.~\eqref{eq:MAE_s1}-\eqref{eq:MAE_s2} to other AM is beyond the scope of this work, and constitute an interesting future project.

\textit{{\red Conclusions.--}} 
In summary, we have applied recently-developed realistic microscopic models for AM to derive relevant Landau coefficients of the free energy, focusing on the coupling between magnetization and Néel order, and magnetic anisotropy energies. The results explain the generic large AHE and the observed strong material-dependence of the SOC-induced weak FM. We stress that the weak FM described in this work refers to the spin moment. For materials where this moment is forbidden to linear order, e.g. the rutiles, secondary orders become relevant, and the orbital magnetic moment may also yield contributions to the small but finite net magnetization~\cite{Smejkal2020Jun,J02024}. Finally, we discovered a general quasi-symmetry enabled by the SOC that is model-independent and allows to determine for which AM symmetries the induced FM spin moment is large or vanishingly small.

\begin{acknowledgments}
\textit{{\red Acknowledgements.--}} 
M.~R. acknow\-ledges support from the Novo Nordisk Foundation grant NNF20OC0060019. A.~K. acknowledges support by the Danish National Committee for Research Infrastructure (NUFI) through the ESS-Lighthouse Q-MAT. D.~F.~A. and Y.~Y. were supported by the National Science Foundation Grant No. DMREF 2323857. Work at UWM  was also supported by a grant from the Simons Foundation (SFI-MPS-NFS-00006741-02, D.F.A and M.R.).

\end{acknowledgments}

\bibliography{Spin_FM_in_AM}

\begin{thebibliography}{44}%
\makeatletter
\providecommand \@ifxundefined [1]{%
 \@ifx{#1\undefined}
}%
\providecommand \@ifnum [1]{%
 \ifnum #1\expandafter \@firstoftwo
 \else \expandafter \@secondoftwo
 \fi
}%
\providecommand \@ifx [1]{%
 \ifx #1\expandafter \@firstoftwo
 \else \expandafter \@secondoftwo
 \fi
}%
\providecommand \natexlab [1]{#1}%
\providecommand \enquote  [1]{``#1''}%
\providecommand \bibnamefont  [1]{#1}%
\providecommand \bibfnamefont [1]{#1}%
\providecommand \citenamefont [1]{#1}%
\providecommand \href@noop [0]{\@secondoftwo}%
\providecommand \href [0]{\begingroup \@sanitize@url \@href}%
\providecommand \@href[1]{\@@startlink{#1}\@@href}%
\providecommand \@@href[1]{\endgroup#1\@@endlink}%
\providecommand \@sanitize@url [0]{\catcode `\\12\catcode `\$12\catcode `\&12\catcode `\#12\catcode `\^12\catcode `\_12\catcode `\%12\relax}%
\providecommand \@@startlink[1]{}%
\providecommand \@@endlink[0]{}%
\providecommand \url  [0]{\begingroup\@sanitize@url \@url }%
\providecommand \@url [1]{\endgroup\@href {#1}{\urlprefix }}%
\providecommand \urlprefix  [0]{URL }%
\providecommand \Eprint [0]{\href }%
\providecommand \doibase [0]{http://dx.doi.org/}%
\providecommand \selectlanguage [0]{\@gobble}%
\providecommand \bibinfo  [0]{\@secondoftwo}%
\providecommand \bibfield  [0]{\@secondoftwo}%
\providecommand \translation [1]{[#1]}%
\providecommand \BibitemOpen [0]{}%
\providecommand \bibitemStop [0]{}%
\providecommand \bibitemNoStop [0]{.\EOS\space}%
\providecommand \EOS [0]{\spacefactor3000\relax}%
\providecommand \BibitemShut  [1]{\csname bibitem#1\endcsname}%
\let\auto@bib@innerbib\@empty
\bibitem [{\citenamefont {Nagaosa}\ \emph {et~al.}(2010)\citenamefont {Nagaosa}, \citenamefont {Sinova}, \citenamefont {Onoda}, \citenamefont {MacDonald},\ and\ \citenamefont {Ong}}]{Nagaosa2010}%
  \BibitemOpen
  \bibfield  {author} {\bibinfo {author} {\bibfnamefont {Naoto}\ \bibnamefont {Nagaosa}}, \bibinfo {author} {\bibfnamefont {Jairo}\ \bibnamefont {Sinova}}, \bibinfo {author} {\bibfnamefont {Shigeki}\ \bibnamefont {Onoda}}, \bibinfo {author} {\bibfnamefont {A.~H.}\ \bibnamefont {MacDonald}}, \ and\ \bibinfo {author} {\bibfnamefont {N.~P.}\ \bibnamefont {Ong}},\ }\bibfield  {title} {\enquote {\bibinfo {title} {{Anomalous Hall effect}},}\ }\href {\doibase 10.1103/RevModPhys.82.1539} {\bibfield  {journal} {\bibinfo  {journal} {Rev. Mod. Phys.}\ }\textbf {\bibinfo {volume} {82}},\ \bibinfo {pages} {1539--1592} (\bibinfo {year} {2010})}\BibitemShut {NoStop}%
\bibitem [{\citenamefont {Sodemann}\ and\ \citenamefont {Fu}(2015)}]{Sodemann2015}%
  \BibitemOpen
  \bibfield  {author} {\bibinfo {author} {\bibfnamefont {Inti}\ \bibnamefont {Sodemann}}\ and\ \bibinfo {author} {\bibfnamefont {Liang}\ \bibnamefont {Fu}},\ }\bibfield  {title} {\enquote {\bibinfo {title} {{Quantum Nonlinear Hall Effect Induced by Berry Curvature Dipole in Time-Reversal Invariant Materials}},}\ }\href {\doibase 10.1103/PhysRevLett.115.216806} {\bibfield  {journal} {\bibinfo  {journal} {Phys. Rev. Lett.}\ }\textbf {\bibinfo {volume} {115}},\ \bibinfo {pages} {216806} (\bibinfo {year} {2015})}\BibitemShut {NoStop}%
\bibitem [{\citenamefont {{Anyuan Gao, Yu-Fei Liu, Jian-Xiang Qiu, Barun Ghosh, Tha{\ifmmode\acute{\imath}\else\'{\i}\fi}s V. Trevisan, Yugo Onishi, Chaowei Hu, Tiema Qian, Hung-Ju Tien, Shao-Wen Chen, Mengqi Huang, Damien B{\ifmmode\acute{e}\else\'{e}\fi}rub{\ifmmode\acute{e}\else\'{e}\fi}, Houchen Li, Christian Tzschaschel, Thao Dinh, \textit{et al.}}}(2023)}]{Gao2023}%
  \BibitemOpen
  \bibfield  {author} {\bibinfo {author} {\bibnamefont {{Anyuan Gao, Yu-Fei Liu, Jian-Xiang Qiu, Barun Ghosh, Tha{\ifmmode\acute{\imath}\else\'{\i}\fi}s V. Trevisan, Yugo Onishi, Chaowei Hu, Tiema Qian, Hung-Ju Tien, Shao-Wen Chen, Mengqi Huang, Damien B{\ifmmode\acute{e}\else\'{e}\fi}rub{\ifmmode\acute{e}\else\'{e}\fi}, Houchen Li, Christian Tzschaschel, Thao Dinh, \textit{et al.}}}},\ }\bibfield  {title} {\enquote {\bibinfo {title} {{Quantum metric nonlinear Hall effect in a topological antiferromagnetic heterostructure}},}\ }\href {\doibase 10.1126/science.adf1506} {\bibfield  {journal} {\bibinfo  {journal} {Science}\ }\textbf {\bibinfo {volume} {381}},\ \bibinfo {pages} {181--186} (\bibinfo {year} {2023})}\BibitemShut {NoStop}%
\bibitem [{\citenamefont {Kaplan}\ \emph {et~al.}(2024)\citenamefont {Kaplan}, \citenamefont {Holder},\ and\ \citenamefont {Yan}}]{Kaplan2024}%
  \BibitemOpen
  \bibfield  {author} {\bibinfo {author} {\bibfnamefont {Daniel}\ \bibnamefont {Kaplan}}, \bibinfo {author} {\bibfnamefont {Tobias}\ \bibnamefont {Holder}}, \ and\ \bibinfo {author} {\bibfnamefont {Binghai}\ \bibnamefont {Yan}},\ }\bibfield  {title} {\enquote {\bibinfo {title} {{Unification of Nonlinear Anomalous Hall Effect and Nonreciprocal Magnetoresistance in Metals by the Quantum Geometry}},}\ }\href {\doibase 10.1103/PhysRevLett.132.026301} {\bibfield  {journal} {\bibinfo  {journal} {Phys. Rev. Lett.}\ }\textbf {\bibinfo {volume} {132}},\ \bibinfo {pages} {026301} (\bibinfo {year} {2024})}\BibitemShut {NoStop}%
\bibitem [{\citenamefont {Fang}\ \emph {et~al.}(2024)\citenamefont {Fang}, \citenamefont {Cano},\ and\ \citenamefont {Ghorashi}}]{Fang2024}%
  \BibitemOpen
  \bibfield  {author} {\bibinfo {author} {\bibfnamefont {Yuan}\ \bibnamefont {Fang}}, \bibinfo {author} {\bibfnamefont {Jennifer}\ \bibnamefont {Cano}}, \ and\ \bibinfo {author} {\bibfnamefont {Sayed Ali~Akbar}\ \bibnamefont {Ghorashi}},\ }\bibfield  {title} {\enquote {\bibinfo {title} {{Quantum Geometry Induced Nonlinear Transport in Altermagnets}},}\ }\href {\doibase 10.1103/PhysRevLett.133.106701} {\bibfield  {journal} {\bibinfo  {journal} {Phys. Rev. Lett.}\ }\textbf {\bibinfo {volume} {133}},\ \bibinfo {pages} {106701} (\bibinfo {year} {2024})}\BibitemShut {NoStop}%
\bibitem [{\citenamefont {Du}\ \emph {et~al.}(2021)\citenamefont {Du}, \citenamefont {Lu},\ and\ \citenamefont {Xie}}]{Du2021}%
  \BibitemOpen
  \bibfield  {author} {\bibinfo {author} {\bibfnamefont {Z.~Z.}\ \bibnamefont {Du}}, \bibinfo {author} {\bibfnamefont {Hai-Zhou}\ \bibnamefont {Lu}}, \ and\ \bibinfo {author} {\bibfnamefont {X.~C.}\ \bibnamefont {Xie}},\ }\bibfield  {title} {\enquote {\bibinfo {title} {{Nonlinear Hall effects}},}\ }\href {\doibase 10.1038/s42254-021-00359-6} {\bibfield  {journal} {\bibinfo  {journal} {Nature Reviews Physics}\ }\textbf {\bibinfo {volume} {3}},\ \bibinfo {pages} {744--752} (\bibinfo {year} {2021})}\BibitemShut {NoStop}%
\bibitem [{\citenamefont {Wang}\ \emph {et~al.}(2023)\citenamefont {Wang}, \citenamefont {Kaplan}, \citenamefont {Zhang}, \citenamefont {Holder}, \citenamefont {Cao}, \citenamefont {Wang}, \citenamefont {Zhou}, \citenamefont {Zhou}, \citenamefont {Jiang}, \citenamefont {Zhang}, \citenamefont {Ru}, \citenamefont {Cai}, \citenamefont {Watanabe}, \citenamefont {Taniguchi}, \citenamefont {Yan},\ and\ \citenamefont {Gao}}]{Wang2023}%
  \BibitemOpen
  \bibfield  {author} {\bibinfo {author} {\bibfnamefont {Naizhou}\ \bibnamefont {Wang}}, \bibinfo {author} {\bibfnamefont {Daniel}\ \bibnamefont {Kaplan}}, \bibinfo {author} {\bibfnamefont {Zhaowei}\ \bibnamefont {Zhang}}, \bibinfo {author} {\bibfnamefont {Tobias}\ \bibnamefont {Holder}}, \bibinfo {author} {\bibfnamefont {Ning}\ \bibnamefont {Cao}}, \bibinfo {author} {\bibfnamefont {Aifeng}\ \bibnamefont {Wang}}, \bibinfo {author} {\bibfnamefont {Xiaoyuan}\ \bibnamefont {Zhou}}, \bibinfo {author} {\bibfnamefont {Feifei}\ \bibnamefont {Zhou}}, \bibinfo {author} {\bibfnamefont {Zhengzhi}\ \bibnamefont {Jiang}}, \bibinfo {author} {\bibfnamefont {Chusheng}\ \bibnamefont {Zhang}}, \bibinfo {author} {\bibfnamefont {Shihao}\ \bibnamefont {Ru}}, \bibinfo {author} {\bibfnamefont {Hongbing}\ \bibnamefont {Cai}}, \bibinfo {author} {\bibfnamefont {Kenji}\ \bibnamefont {Watanabe}}, \bibinfo {author} {\bibfnamefont {Takashi}\ \bibnamefont {Taniguchi}}, \bibinfo {author} {\bibfnamefont {Binghai}\ \bibnamefont {Yan}}, \ and\
  \bibinfo {author} {\bibfnamefont {Weibo}\ \bibnamefont {Gao}},\ }\bibfield  {title} {\enquote {\bibinfo {title} {{Quantum-metric-induced nonlinear transport in a topological antiferromagnet}},}\ }\href {\doibase 10.1038/s41586-023-06363-3} {\bibfield  {journal} {\bibinfo  {journal} {Nature}\ }\textbf {\bibinfo {volume} {621}},\ \bibinfo {pages} {487--492} (\bibinfo {year} {2023})}\BibitemShut {NoStop}%
\bibitem [{\citenamefont {Lee}\ \emph {et~al.}(2024)\citenamefont {Lee}, \citenamefont {Lee}, \citenamefont {Jung}, \citenamefont {Jung}, \citenamefont {Kim}, \citenamefont {Lee}, \citenamefont {Seok}, \citenamefont {Kim}, \citenamefont {Park}, \citenamefont {{\ifmmode\check{S}\else\v{S}\fi}mejkal}, \citenamefont {Kang},\ and\ \citenamefont {Kim}}]{Lee2024Jan}%
  \BibitemOpen
  \bibfield  {author} {\bibinfo {author} {\bibfnamefont {Suyoung}\ \bibnamefont {Lee}}, \bibinfo {author} {\bibfnamefont {Sangjae}\ \bibnamefont {Lee}}, \bibinfo {author} {\bibfnamefont {Saegyeol}\ \bibnamefont {Jung}}, \bibinfo {author} {\bibfnamefont {Jiwon}\ \bibnamefont {Jung}}, \bibinfo {author} {\bibfnamefont {Donghan}\ \bibnamefont {Kim}}, \bibinfo {author} {\bibfnamefont {Yeonjae}\ \bibnamefont {Lee}}, \bibinfo {author} {\bibfnamefont {Byeongjun}\ \bibnamefont {Seok}}, \bibinfo {author} {\bibfnamefont {Jaeyoung}\ \bibnamefont {Kim}}, \bibinfo {author} {\bibfnamefont {Byeong~Gyu}\ \bibnamefont {Park}}, \bibinfo {author} {\bibfnamefont {Libor}\ \bibnamefont {{\ifmmode\check{S}\else\v{S}\fi}mejkal}}, \bibinfo {author} {\bibfnamefont {Chang-Jong}\ \bibnamefont {Kang}}, \ and\ \bibinfo {author} {\bibfnamefont {Changyoung}\ \bibnamefont {Kim}},\ }\bibfield  {title} {\enquote {\bibinfo {title} {{Broken Kramers Degeneracy in Altermagnetic MnTe}},}\ }\href {\doibase 10.1103/PhysRevLett.132.036702} {\bibfield
  {journal} {\bibinfo  {journal} {Phys. Rev. Lett.}\ }\textbf {\bibinfo {volume} {132}},\ \bibinfo {pages} {036702} (\bibinfo {year} {2024})}\BibitemShut {NoStop}%
\bibitem [{\citenamefont {Krempask{\ifmmode\acute{y}\else\'{y}\fi}}\ \emph {et~al.}(2024)\citenamefont {Krempask{\ifmmode\acute{y}\else\'{y}\fi}}, \citenamefont {{\ifmmode\check{S}\else\v{S}\fi}mejkal}, \citenamefont {D{'}Souza}, \citenamefont {Hajlaoui}, \citenamefont {Springholz}, \citenamefont {Uhl{\ifmmode\acute{\imath}\else\'{\i}\fi}{\ifmmode\check{r}\else\v{r}\fi}ov{\ifmmode\acute{a}\else\'{a}\fi}}, \citenamefont {Alarab}, \citenamefont {Constantinou}, \citenamefont {Strocov}, \citenamefont {Usanov}, \citenamefont {Pudelko}, \citenamefont {Gonz{\ifmmode\acute{a}\else\'{a}\fi}lez-Hern{\ifmmode\acute{a}\else\'{a}\fi}ndez}, \citenamefont {Birk~Hellenes}, \citenamefont {Jansa}, \citenamefont {Reichlov{\ifmmode\acute{a}\else\'{a}\fi}}, \citenamefont {{\ifmmode\check{S}\else\v{S}\fi}ob{\ifmmode\acute{a}\else\'{a}\fi}{\ifmmode\check{n}\else\v{n}\fi}}, \citenamefont {Gonzalez~Betancourt}, \citenamefont {Wadley}, \citenamefont {Sinova}, \citenamefont {Kriegner}, \citenamefont
  {Min{\ifmmode\acute{a}\else\'{a}\fi}r}, \citenamefont {Dil},\ and\ \citenamefont {Jungwirth}}]{Krempasky2024Feb}%
  \BibitemOpen
  \bibfield  {author} {\bibinfo {author} {\bibfnamefont {J.}~\bibnamefont {Krempask{\ifmmode\acute{y}\else\'{y}\fi}}}, \bibinfo {author} {\bibfnamefont {L.}~\bibnamefont {{\ifmmode\check{S}\else\v{S}\fi}mejkal}}, \bibinfo {author} {\bibfnamefont {S.~W.}\ \bibnamefont {D{'}Souza}}, \bibinfo {author} {\bibfnamefont {M.}~\bibnamefont {Hajlaoui}}, \bibinfo {author} {\bibfnamefont {G.}~\bibnamefont {Springholz}}, \bibinfo {author} {\bibfnamefont {K.}~\bibnamefont {Uhl{\ifmmode\acute{\imath}\else\'{\i}\fi}{\ifmmode\check{r}\else\v{r}\fi}ov{\ifmmode\acute{a}\else\'{a}\fi}}}, \bibinfo {author} {\bibfnamefont {F.}~\bibnamefont {Alarab}}, \bibinfo {author} {\bibfnamefont {P.~C.}\ \bibnamefont {Constantinou}}, \bibinfo {author} {\bibfnamefont {V.}~\bibnamefont {Strocov}}, \bibinfo {author} {\bibfnamefont {D.}~\bibnamefont {Usanov}}, \bibinfo {author} {\bibfnamefont {W.~R.}\ \bibnamefont {Pudelko}}, \bibinfo {author} {\bibfnamefont {R.}~\bibnamefont
  {Gonz{\ifmmode\acute{a}\else\'{a}\fi}lez-Hern{\ifmmode\acute{a}\else\'{a}\fi}ndez}}, \bibinfo {author} {\bibfnamefont {A.}~\bibnamefont {Birk~Hellenes}}, \bibinfo {author} {\bibfnamefont {Z.}~\bibnamefont {Jansa}}, \bibinfo {author} {\bibfnamefont {H.}~\bibnamefont {Reichlov{\ifmmode\acute{a}\else\'{a}\fi}}}, \bibinfo {author} {\bibfnamefont {Z.}~\bibnamefont {{\ifmmode\check{S}\else\v{S}\fi}ob{\ifmmode\acute{a}\else\'{a}\fi}{\ifmmode\check{n}\else\v{n}\fi}}}, \bibinfo {author} {\bibfnamefont {R.~D.}\ \bibnamefont {Gonzalez~Betancourt}}, \bibinfo {author} {\bibfnamefont {P.}~\bibnamefont {Wadley}}, \bibinfo {author} {\bibfnamefont {J.}~\bibnamefont {Sinova}}, \bibinfo {author} {\bibfnamefont {D.}~\bibnamefont {Kriegner}}, \bibinfo {author} {\bibfnamefont {J.}~\bibnamefont {Min{\ifmmode\acute{a}\else\'{a}\fi}r}}, \bibinfo {author} {\bibfnamefont {J.~H.}\ \bibnamefont {Dil}}, \ and\ \bibinfo {author} {\bibfnamefont {T.}~\bibnamefont {Jungwirth}},\ }\bibfield  {title} {\enquote {\bibinfo {title}
  {{Altermagnetic lifting of Kramers spin degeneracy}},}\ }\href {\doibase 10.1038/s41586-023-06907-7} {\bibfield  {journal} {\bibinfo  {journal} {Nature}\ }\textbf {\bibinfo {volume} {626}},\ \bibinfo {pages} {517--522} (\bibinfo {year} {2024})}\BibitemShut {NoStop}%
\bibitem [{\citenamefont {Osumi}\ \emph {et~al.}(2024)\citenamefont {Osumi}, \citenamefont {Souma}, \citenamefont {Aoyama}, \citenamefont {Yamauchi}, \citenamefont {Honma}, \citenamefont {Nakayama}, \citenamefont {Takahashi}, \citenamefont {Ohgushi},\ and\ \citenamefont {Sato}}]{Osumi2024Mar}%
  \BibitemOpen
  \bibfield  {author} {\bibinfo {author} {\bibfnamefont {T.}~\bibnamefont {Osumi}}, \bibinfo {author} {\bibfnamefont {S.}~\bibnamefont {Souma}}, \bibinfo {author} {\bibfnamefont {T.}~\bibnamefont {Aoyama}}, \bibinfo {author} {\bibfnamefont {K.}~\bibnamefont {Yamauchi}}, \bibinfo {author} {\bibfnamefont {A.}~\bibnamefont {Honma}}, \bibinfo {author} {\bibfnamefont {K.}~\bibnamefont {Nakayama}}, \bibinfo {author} {\bibfnamefont {T.}~\bibnamefont {Takahashi}}, \bibinfo {author} {\bibfnamefont {K.}~\bibnamefont {Ohgushi}}, \ and\ \bibinfo {author} {\bibfnamefont {T.}~\bibnamefont {Sato}},\ }\bibfield  {title} {\enquote {\bibinfo {title} {{Observation of a giant band splitting in altermagnetic MnTe}},}\ }\href {\doibase 10.1103/PhysRevB.109.115102} {\bibfield  {journal} {\bibinfo  {journal} {Phys. Rev. B}\ }\textbf {\bibinfo {volume} {109}},\ \bibinfo {pages} {115102} (\bibinfo {year} {2024})}\BibitemShut {NoStop}%
\bibitem [{\citenamefont {Li}\ \emph {et~al.}(2024{\natexlab{a}})\citenamefont {Li}, \citenamefont {Hu}, \citenamefont {Li}, \citenamefont {Wang}, \citenamefont {Chen}, \citenamefont {Thiagarajan}, \citenamefont {Leandersson}, \citenamefont {Polley}, \citenamefont {Kim}, \citenamefont {Liu}, \citenamefont {Fulga}, \citenamefont {Vergniory}, \citenamefont {Janson}, \citenamefont {Tjernberg},\ and\ \citenamefont {Brink}}]{Li2024May}%
  \BibitemOpen
  \bibfield  {author} {\bibinfo {author} {\bibfnamefont {Cong}\ \bibnamefont {Li}}, \bibinfo {author} {\bibfnamefont {Mengli}\ \bibnamefont {Hu}}, \bibinfo {author} {\bibfnamefont {Zhilin}\ \bibnamefont {Li}}, \bibinfo {author} {\bibfnamefont {Yang}\ \bibnamefont {Wang}}, \bibinfo {author} {\bibfnamefont {Wanyu}\ \bibnamefont {Chen}}, \bibinfo {author} {\bibfnamefont {Balasubramanian}\ \bibnamefont {Thiagarajan}}, \bibinfo {author} {\bibfnamefont {Mats}\ \bibnamefont {Leandersson}}, \bibinfo {author} {\bibfnamefont {Craig}\ \bibnamefont {Polley}}, \bibinfo {author} {\bibfnamefont {Timur}\ \bibnamefont {Kim}}, \bibinfo {author} {\bibfnamefont {Hui}\ \bibnamefont {Liu}}, \bibinfo {author} {\bibfnamefont {Cosma}\ \bibnamefont {Fulga}}, \bibinfo {author} {\bibfnamefont {Maia~G.}\ \bibnamefont {Vergniory}}, \bibinfo {author} {\bibfnamefont {Oleg}\ \bibnamefont {Janson}}, \bibinfo {author} {\bibfnamefont {Oscar}\ \bibnamefont {Tjernberg}}, \ and\ \bibinfo {author} {\bibfnamefont {Jeroen van~den}\ \bibnamefont
  {Brink}},\ }\bibfield  {title} {\enquote {\bibinfo {title} {{Topological Weyl Altermagnetism in CrSb}},}\ }\href {\doibase 10.48550/arXiv.2405.14777} {\bibfield  {journal} {\bibinfo  {journal} {arXiv}\ } (\bibinfo {year} {2024}{\natexlab{a}}),\ 10.48550/arXiv.2405.14777},\ \Eprint {http://arxiv.org/abs/2405.14777} {2405.14777} \BibitemShut {NoStop}%
\bibitem [{\citenamefont {Yang}\ \emph {et~al.}(2025)\citenamefont {Yang}, \citenamefont {Li}, \citenamefont {Yang}, \citenamefont {Li}, \citenamefont {Zheng}, \citenamefont {Zhu}, \citenamefont {Pan}, \citenamefont {Xu}, \citenamefont {Cao}, \citenamefont {Zhao}, \citenamefont {Jana}, \citenamefont {Zhang}, \citenamefont {Ye}, \citenamefont {Song}, \citenamefont {Hu}, \citenamefont {Yang}, \citenamefont {Fujii}, \citenamefont {Vobornik}, \citenamefont {Shi}, \citenamefont {Yuan}, \citenamefont {Zhang}, \citenamefont {Xu},\ and\ \citenamefont {Liu}}]{Yang2024May}%
  \BibitemOpen
  \bibfield  {author} {\bibinfo {author} {\bibfnamefont {Guowei}\ \bibnamefont {Yang}}, \bibinfo {author} {\bibfnamefont {Zhanghuan}\ \bibnamefont {Li}}, \bibinfo {author} {\bibfnamefont {Sai}\ \bibnamefont {Yang}}, \bibinfo {author} {\bibfnamefont {Jiyuan}\ \bibnamefont {Li}}, \bibinfo {author} {\bibfnamefont {Hao}\ \bibnamefont {Zheng}}, \bibinfo {author} {\bibfnamefont {Weifan}\ \bibnamefont {Zhu}}, \bibinfo {author} {\bibfnamefont {Ze}~\bibnamefont {Pan}}, \bibinfo {author} {\bibfnamefont {Yifu}\ \bibnamefont {Xu}}, \bibinfo {author} {\bibfnamefont {Saizheng}\ \bibnamefont {Cao}}, \bibinfo {author} {\bibfnamefont {Wenxuan}\ \bibnamefont {Zhao}}, \bibinfo {author} {\bibfnamefont {Anupam}\ \bibnamefont {Jana}}, \bibinfo {author} {\bibfnamefont {Jiawen}\ \bibnamefont {Zhang}}, \bibinfo {author} {\bibfnamefont {Mao}\ \bibnamefont {Ye}}, \bibinfo {author} {\bibfnamefont {Yu}~\bibnamefont {Song}}, \bibinfo {author} {\bibfnamefont {Lun-Hui}\ \bibnamefont {Hu}}, \bibinfo {author} {\bibfnamefont {Lexian}\
  \bibnamefont {Yang}}, \bibinfo {author} {\bibfnamefont {Jun}\ \bibnamefont {Fujii}}, \bibinfo {author} {\bibfnamefont {Ivana}\ \bibnamefont {Vobornik}}, \bibinfo {author} {\bibfnamefont {Ming}\ \bibnamefont {Shi}}, \bibinfo {author} {\bibfnamefont {Huiqiu}\ \bibnamefont {Yuan}}, \bibinfo {author} {\bibfnamefont {Yongjun}\ \bibnamefont {Zhang}}, \bibinfo {author} {\bibfnamefont {Yuanfeng}\ \bibnamefont {Xu}}, \ and\ \bibinfo {author} {\bibfnamefont {Yang}\ \bibnamefont {Liu}},\ }\bibfield  {title} {\enquote {\bibinfo {title} {{Three-dimensional mapping of the altermagnetic spin splitting in CrSb}},}\ }\href {\doibase 10.1038/s41467-025-56647-7} {\bibfield  {journal} {\bibinfo  {journal} {Nature Communications}\ }\textbf {\bibinfo {volume} {16}},\ \bibinfo {pages} {1442} (\bibinfo {year} {2025})}\BibitemShut {NoStop}%
\bibitem [{\citenamefont {Reimers}\ \emph {et~al.}(2024)\citenamefont {Reimers}, \citenamefont {Odenbreit}, \citenamefont {{\ifmmode\check{S}\else\v{S}\fi}mejkal}, \citenamefont {Strocov}, \citenamefont {Constantinou}, \citenamefont {Hellenes}, \citenamefont {Jaeschke~Ubiergo}, \citenamefont {Campos}, \citenamefont {Bharadwaj}, \citenamefont {Chakraborty}, \citenamefont {Denneulin}, \citenamefont {Shi}, \citenamefont {Dunin-Borkowski}, \citenamefont {Das}, \citenamefont {Kl{\ifmmode\ddot{a}\else\"{a}\fi}ui}, \citenamefont {Sinova},\ and\ \citenamefont {Jourdan}}]{Reimers2024Mar}%
  \BibitemOpen
  \bibfield  {author} {\bibinfo {author} {\bibfnamefont {Sonka}\ \bibnamefont {Reimers}}, \bibinfo {author} {\bibfnamefont {Lukas}\ \bibnamefont {Odenbreit}}, \bibinfo {author} {\bibfnamefont {Libor}\ \bibnamefont {{\ifmmode\check{S}\else\v{S}\fi}mejkal}}, \bibinfo {author} {\bibfnamefont {Vladimir~N.}\ \bibnamefont {Strocov}}, \bibinfo {author} {\bibfnamefont {Procopios}\ \bibnamefont {Constantinou}}, \bibinfo {author} {\bibfnamefont {Anna~B.}\ \bibnamefont {Hellenes}}, \bibinfo {author} {\bibfnamefont {Rodrigo}\ \bibnamefont {Jaeschke~Ubiergo}}, \bibinfo {author} {\bibfnamefont {Warlley~H.}\ \bibnamefont {Campos}}, \bibinfo {author} {\bibfnamefont {Venkata~K.}\ \bibnamefont {Bharadwaj}}, \bibinfo {author} {\bibfnamefont {Atasi}\ \bibnamefont {Chakraborty}}, \bibinfo {author} {\bibfnamefont {Thibaud}\ \bibnamefont {Denneulin}}, \bibinfo {author} {\bibfnamefont {Wen}\ \bibnamefont {Shi}}, \bibinfo {author} {\bibfnamefont {Rafal~E.}\ \bibnamefont {Dunin-Borkowski}}, \bibinfo {author} {\bibfnamefont {Suvadip}\
  \bibnamefont {Das}}, \bibinfo {author} {\bibfnamefont {Mathias}\ \bibnamefont {Kl{\ifmmode\ddot{a}\else\"{a}\fi}ui}}, \bibinfo {author} {\bibfnamefont {Jairo}\ \bibnamefont {Sinova}}, \ and\ \bibinfo {author} {\bibfnamefont {Martin}\ \bibnamefont {Jourdan}},\ }\bibfield  {title} {\enquote {\bibinfo {title} {{Direct observation of altermagnetic band splitting in CrSb thin films}},}\ }\href {\doibase 10.1038/s41467-024-46476-5} {\bibfield  {journal} {\bibinfo  {journal} {Nat. Commun.}\ }\textbf {\bibinfo {volume} {15}},\ \bibinfo {pages} {1--7} (\bibinfo {year} {2024})}\BibitemShut {NoStop}%
\bibitem [{\citenamefont {Ding}\ \emph {et~al.}(2024)\citenamefont {Ding}, \citenamefont {Jiang}, \citenamefont {Chen}, \citenamefont {Tao}, \citenamefont {Liu}, \citenamefont {Li}, \citenamefont {Liu}, \citenamefont {Sun}, \citenamefont {Cheng}, \citenamefont {Liu}, \citenamefont {Yang}, \citenamefont {Zhang}, \citenamefont {Deng}, \citenamefont {Jing}, \citenamefont {Huang}, \citenamefont {Shi}, \citenamefont {Ye}, \citenamefont {Qiao}, \citenamefont {Wang}, \citenamefont {Guo}, \citenamefont {Feng},\ and\ \citenamefont {Shen}}]{Ding2024}%
  \BibitemOpen
  \bibfield  {author} {\bibinfo {author} {\bibfnamefont {Jianyang}\ \bibnamefont {Ding}}, \bibinfo {author} {\bibfnamefont {Zhicheng}\ \bibnamefont {Jiang}}, \bibinfo {author} {\bibfnamefont {Xiuhua}\ \bibnamefont {Chen}}, \bibinfo {author} {\bibfnamefont {Zicheng}\ \bibnamefont {Tao}}, \bibinfo {author} {\bibfnamefont {Zhengtai}\ \bibnamefont {Liu}}, \bibinfo {author} {\bibfnamefont {Tongrui}\ \bibnamefont {Li}}, \bibinfo {author} {\bibfnamefont {Jishan}\ \bibnamefont {Liu}}, \bibinfo {author} {\bibfnamefont {Jianping}\ \bibnamefont {Sun}}, \bibinfo {author} {\bibfnamefont {Jinguang}\ \bibnamefont {Cheng}}, \bibinfo {author} {\bibfnamefont {Jiayu}\ \bibnamefont {Liu}}, \bibinfo {author} {\bibfnamefont {Yichen}\ \bibnamefont {Yang}}, \bibinfo {author} {\bibfnamefont {Runfeng}\ \bibnamefont {Zhang}}, \bibinfo {author} {\bibfnamefont {Liwei}\ \bibnamefont {Deng}}, \bibinfo {author} {\bibfnamefont {Wenchuan}\ \bibnamefont {Jing}}, \bibinfo {author} {\bibfnamefont {Yu}~\bibnamefont {Huang}}, \bibinfo {author}
  {\bibfnamefont {Yuming}\ \bibnamefont {Shi}}, \bibinfo {author} {\bibfnamefont {Mao}\ \bibnamefont {Ye}}, \bibinfo {author} {\bibfnamefont {Shan}\ \bibnamefont {Qiao}}, \bibinfo {author} {\bibfnamefont {Yilin}\ \bibnamefont {Wang}}, \bibinfo {author} {\bibfnamefont {Yanfeng}\ \bibnamefont {Guo}}, \bibinfo {author} {\bibfnamefont {Donglai}\ \bibnamefont {Feng}}, \ and\ \bibinfo {author} {\bibfnamefont {Dawei}\ \bibnamefont {Shen}},\ }\bibfield  {title} {\enquote {\bibinfo {title} {{Large Band Splitting in $g$-Wave Altermagnet CrSb}},}\ }\href {\doibase 10.1103/PhysRevLett.133.206401} {\bibfield  {journal} {\bibinfo  {journal} {Phys. Rev. Lett.}\ }\textbf {\bibinfo {volume} {133}},\ \bibinfo {pages} {206401} (\bibinfo {year} {2024})}\BibitemShut {NoStop}%
\bibitem [{\citenamefont {Feng}\ \emph {et~al.}(2022)\citenamefont {Feng}, \citenamefont {Zhou}, \citenamefont {{\ifmmode\check{S}\else\v{S}\fi}mejkal}, \citenamefont {Wu}, \citenamefont {Zhu}, \citenamefont {Guo}, \citenamefont {Gonz{\ifmmode\acute{a}\else\'{a}\fi}lez-Hern{\ifmmode\acute{a}\else\'{a}\fi}ndez}, \citenamefont {Wang}, \citenamefont {Yan}, \citenamefont {Qin}, \citenamefont {Zhang}, \citenamefont {Wu}, \citenamefont {Chen}, \citenamefont {Meng}, \citenamefont {Liu}, \citenamefont {Xia}, \citenamefont {Sinova}, \citenamefont {Jungwirth},\ and\ \citenamefont {Liu}}]{Feng2022Nov}%
  \BibitemOpen
  \bibfield  {author} {\bibinfo {author} {\bibfnamefont {Zexin}\ \bibnamefont {Feng}}, \bibinfo {author} {\bibfnamefont {Xiaorong}\ \bibnamefont {Zhou}}, \bibinfo {author} {\bibfnamefont {Libor}\ \bibnamefont {{\ifmmode\check{S}\else\v{S}\fi}mejkal}}, \bibinfo {author} {\bibfnamefont {Lei}\ \bibnamefont {Wu}}, \bibinfo {author} {\bibfnamefont {Zengwei}\ \bibnamefont {Zhu}}, \bibinfo {author} {\bibfnamefont {Huixin}\ \bibnamefont {Guo}}, \bibinfo {author} {\bibfnamefont {Rafael}\ \bibnamefont {Gonz{\ifmmode\acute{a}\else\'{a}\fi}lez-Hern{\ifmmode\acute{a}\else\'{a}\fi}ndez}}, \bibinfo {author} {\bibfnamefont {Xiaoning}\ \bibnamefont {Wang}}, \bibinfo {author} {\bibfnamefont {Han}\ \bibnamefont {Yan}}, \bibinfo {author} {\bibfnamefont {Peixin}\ \bibnamefont {Qin}}, \bibinfo {author} {\bibfnamefont {Xin}\ \bibnamefont {Zhang}}, \bibinfo {author} {\bibfnamefont {Haojiang}\ \bibnamefont {Wu}}, \bibinfo {author} {\bibfnamefont {Hongyu}\ \bibnamefont {Chen}}, \bibinfo {author} {\bibfnamefont {Ziang}\ \bibnamefont
  {Meng}}, \bibinfo {author} {\bibfnamefont {Li}~\bibnamefont {Liu}}, \bibinfo {author} {\bibfnamefont {Zhengcai}\ \bibnamefont {Xia}}, \bibinfo {author} {\bibfnamefont {Jairo}\ \bibnamefont {Sinova}}, \bibinfo {author} {\bibfnamefont {Tom{\ifmmode\acute{a}\else\'{a}\fi}{\ifmmode\check{s}\else\v{s}\fi}}\ \bibnamefont {Jungwirth}}, \ and\ \bibinfo {author} {\bibfnamefont {Zhiqi}\ \bibnamefont {Liu}},\ }\bibfield  {title} {\enquote {\bibinfo {title} {{An anomalous Hall effect in altermagnetic ruthenium dioxide}},}\ }\href {\doibase 10.1038/s41928-022-00866-z} {\bibfield  {journal} {\bibinfo  {journal} {Nat. Electron.}\ }\textbf {\bibinfo {volume} {5}},\ \bibinfo {pages} {735--743} (\bibinfo {year} {2022})}\BibitemShut {NoStop}%
\bibitem [{\citenamefont {Kluczyk}\ \emph {et~al.}(2024)\citenamefont {Kluczyk}, \citenamefont {Gas}, \citenamefont {Grzybowski}, \citenamefont {Skupiński}, \citenamefont {Borysiewicz}, \citenamefont {Fas}, \citenamefont {Suffczyński}, \citenamefont {Domagala}, \citenamefont {Grasza}, \citenamefont {Mycielski}, \citenamefont {Baj}, \citenamefont {Ahn}, \citenamefont {Výborný}, \citenamefont {Sawicki},\ and\ \citenamefont {Gryglas-Borysiewicz}}]{Kluczyk2023Oct}%
  \BibitemOpen
  \bibfield  {author} {\bibinfo {author} {\bibfnamefont {K.~P.}\ \bibnamefont {Kluczyk}}, \bibinfo {author} {\bibfnamefont {K.}~\bibnamefont {Gas}}, \bibinfo {author} {\bibfnamefont {M.~J.}\ \bibnamefont {Grzybowski}}, \bibinfo {author} {\bibfnamefont {P.}~\bibnamefont {Skupiński}}, \bibinfo {author} {\bibfnamefont {M.~A.}\ \bibnamefont {Borysiewicz}}, \bibinfo {author} {\bibfnamefont {T.}~\bibnamefont {Fas}}, \bibinfo {author} {\bibfnamefont {J.}~\bibnamefont {Suffczyński}}, \bibinfo {author} {\bibfnamefont {J.~Z.}\ \bibnamefont {Domagala}}, \bibinfo {author} {\bibfnamefont {K.}~\bibnamefont {Grasza}}, \bibinfo {author} {\bibfnamefont {A.}~\bibnamefont {Mycielski}}, \bibinfo {author} {\bibfnamefont {M.}~\bibnamefont {Baj}}, \bibinfo {author} {\bibfnamefont {K.~H.}\ \bibnamefont {Ahn}}, \bibinfo {author} {\bibfnamefont {K.}~\bibnamefont {Výborný}}, \bibinfo {author} {\bibfnamefont {M.}~\bibnamefont {Sawicki}}, \ and\ \bibinfo {author} {\bibfnamefont {M.}~\bibnamefont {Gryglas-Borysiewicz}},\ }\bibfield
  {title} {\enquote {\bibinfo {title} {{Coexistence of anomalous Hall effect and weak magnetization in a nominally collinear antiferromagnet MnTe}},}\ }\href {\doibase 10.1103/PhysRevB.110.155201} {\bibfield  {journal} {\bibinfo  {journal} {Phys. Rev. B}\ }\textbf {\bibinfo {volume} {110}},\ \bibinfo {pages} {155201} (\bibinfo {year} {2024})}\BibitemShut {NoStop}%
\bibitem [{\citenamefont {Gonzalez~Betancourt}\ \emph {et~al.}(2023)\citenamefont {Gonzalez~Betancourt}, \citenamefont {Zub{\ifmmode\acute{a}\else\'{a}\fi}{\ifmmode\check{c}\else\v{c}\fi}}, \citenamefont {Gonzalez-Hernandez}, \citenamefont {Geishendorf}, \citenamefont {{\ifmmode\check{S}\else\v{S}\fi}ob{\ifmmode\acute{a}\else\'{a}\fi}{\ifmmode\check{n}\else\v{n}\fi}}, \citenamefont {Springholz}, \citenamefont {Olejn{\ifmmode\acute{\imath}\else\'{\i}\fi}k}, \citenamefont {{\ifmmode\check{S}\else\v{S}\fi}mejkal}, \citenamefont {Sinova}, \citenamefont {Jungwirth}, \citenamefont {Goennenwein}, \citenamefont {Thomas}, \citenamefont {Reichlov{\ifmmode\acute{a}\else\'{a}\fi}}, \citenamefont {{\ifmmode\check{Z}\else\v{Z}\fi}elezn{\ifmmode\acute{y}\else\'{y}\fi}},\ and\ \citenamefont {Kriegner}}]{GonzalezBetancourt2023Jan}%
  \BibitemOpen
  \bibfield  {author} {\bibinfo {author} {\bibfnamefont {R.~D.}\ \bibnamefont {Gonzalez~Betancourt}}, \bibinfo {author} {\bibfnamefont {J.}~\bibnamefont {Zub{\ifmmode\acute{a}\else\'{a}\fi}{\ifmmode\check{c}\else\v{c}\fi}}}, \bibinfo {author} {\bibfnamefont {R.}~\bibnamefont {Gonzalez-Hernandez}}, \bibinfo {author} {\bibfnamefont {K.}~\bibnamefont {Geishendorf}}, \bibinfo {author} {\bibfnamefont {Z.}~\bibnamefont {{\ifmmode\check{S}\else\v{S}\fi}ob{\ifmmode\acute{a}\else\'{a}\fi}{\ifmmode\check{n}\else\v{n}\fi}}}, \bibinfo {author} {\bibfnamefont {G.}~\bibnamefont {Springholz}}, \bibinfo {author} {\bibfnamefont {K.}~\bibnamefont {Olejn{\ifmmode\acute{\imath}\else\'{\i}\fi}k}}, \bibinfo {author} {\bibfnamefont {L.}~\bibnamefont {{\ifmmode\check{S}\else\v{S}\fi}mejkal}}, \bibinfo {author} {\bibfnamefont {J.}~\bibnamefont {Sinova}}, \bibinfo {author} {\bibfnamefont {T.}~\bibnamefont {Jungwirth}}, \bibinfo {author} {\bibfnamefont {S.~T.~B.}\ \bibnamefont {Goennenwein}}, \bibinfo {author} {\bibfnamefont
  {A.}~\bibnamefont {Thomas}}, \bibinfo {author} {\bibfnamefont {H.}~\bibnamefont {Reichlov{\ifmmode\acute{a}\else\'{a}\fi}}}, \bibinfo {author} {\bibfnamefont {J.}~\bibnamefont {{\ifmmode\check{Z}\else\v{Z}\fi}elezn{\ifmmode\acute{y}\else\'{y}\fi}}}, \ and\ \bibinfo {author} {\bibfnamefont {D.}~\bibnamefont {Kriegner}},\ }\bibfield  {title} {\enquote {\bibinfo {title} {{Spontaneous Anomalous Hall Effect Arising from an Unconventional Compensated Magnetic Phase in a Semiconductor}},}\ }\href {\doibase 10.1103/PhysRevLett.130.036702} {\bibfield  {journal} {\bibinfo  {journal} {Phys. Rev. Lett.}\ }\textbf {\bibinfo {volume} {130}},\ \bibinfo {pages} {036702} (\bibinfo {year} {2023})}\BibitemShut {NoStop}%
\bibitem [{\citenamefont {{\ifmmode\check{S}\else\v{S}\fi}mejkal}\ \emph {et~al.}(2020)\citenamefont {{\ifmmode\check{S}\else\v{S}\fi}mejkal}, \citenamefont {Gonz{\ifmmode\acute{a}\else\'{a}\fi}lez-Hern{\ifmmode\acute{a}\else\'{a}\fi}ndez}, \citenamefont {Jungwirth},\ and\ \citenamefont {Sinova}}]{Smejkal2020Jun}%
  \BibitemOpen
  \bibfield  {author} {\bibinfo {author} {\bibfnamefont {Libor}\ \bibnamefont {{\ifmmode\check{S}\else\v{S}\fi}mejkal}}, \bibinfo {author} {\bibfnamefont {Rafael}\ \bibnamefont {Gonz{\ifmmode\acute{a}\else\'{a}\fi}lez-Hern{\ifmmode\acute{a}\else\'{a}\fi}ndez}}, \bibinfo {author} {\bibfnamefont {T.}~\bibnamefont {Jungwirth}}, \ and\ \bibinfo {author} {\bibfnamefont {J.}~\bibnamefont {Sinova}},\ }\bibfield  {title} {\enquote {\bibinfo {title} {{Crystal time-reversal symmetry breaking and spontaneous Hall effect in collinear antiferromagnets}},}\ }\href {\doibase 10.1126/sciadv.aaz8809} {\bibfield  {journal} {\bibinfo  {journal} {Sci. Adv.}\ }\textbf {\bibinfo {volume} {6}} (\bibinfo {year} {2020}),\ 10.1126/sciadv.aaz8809}\BibitemShut {NoStop}%
\bibitem [{\citenamefont {{\ifmmode\check{S}\else\v{S}\fi}mejkal}\ \emph {et~al.}(2022{\natexlab{a}})\citenamefont {{\ifmmode\check{S}\else\v{S}\fi}mejkal}, \citenamefont {Sinova},\ and\ \citenamefont {Jungwirth}}]{Smejkal2022Sep}%
  \BibitemOpen
  \bibfield  {author} {\bibinfo {author} {\bibfnamefont {Libor}\ \bibnamefont {{\ifmmode\check{S}\else\v{S}\fi}mejkal}}, \bibinfo {author} {\bibfnamefont {Jairo}\ \bibnamefont {Sinova}}, \ and\ \bibinfo {author} {\bibfnamefont {Tomas}\ \bibnamefont {Jungwirth}},\ }\bibfield  {title} {\enquote {\bibinfo {title} {{Beyond Conventional Ferromagnetism and Antiferromagnetism: A Phase with Nonrelativistic Spin and Crystal Rotation Symmetry}},}\ }\href {\doibase 10.1103/PhysRevX.12.031042} {\bibfield  {journal} {\bibinfo  {journal} {Phys. Rev. X}\ }\textbf {\bibinfo {volume} {12}},\ \bibinfo {pages} {031042} (\bibinfo {year} {2022}{\natexlab{a}})}\BibitemShut {NoStop}%
\bibitem [{\citenamefont {{\ifmmode\check{S}\else\v{S}\fi}mejkal}\ \emph {et~al.}(2022{\natexlab{b}})\citenamefont {{\ifmmode\check{S}\else\v{S}\fi}mejkal}, \citenamefont {Sinova},\ and\ \citenamefont {Jungwirth}}]{Smejkal2022Dec}%
  \BibitemOpen
  \bibfield  {author} {\bibinfo {author} {\bibfnamefont {Libor}\ \bibnamefont {{\ifmmode\check{S}\else\v{S}\fi}mejkal}}, \bibinfo {author} {\bibfnamefont {Jairo}\ \bibnamefont {Sinova}}, \ and\ \bibinfo {author} {\bibfnamefont {Tomas}\ \bibnamefont {Jungwirth}},\ }\bibfield  {title} {\enquote {\bibinfo {title} {{Emerging Research Landscape of Altermagnetism}},}\ }\href {\doibase 10.1103/PhysRevX.12.040501} {\bibfield  {journal} {\bibinfo  {journal} {Phys. Rev. X}\ }\textbf {\bibinfo {volume} {12}},\ \bibinfo {pages} {040501} (\bibinfo {year} {2022}{\natexlab{b}})}\BibitemShut {NoStop}%
\bibitem [{\citenamefont {{\ifmmode\check{S}\else\v{S}\fi}mejkal}\ \emph {et~al.}(2022{\natexlab{c}})\citenamefont {{\ifmmode\check{S}\else\v{S}\fi}mejkal}, \citenamefont {MacDonald}, \citenamefont {Sinova}, \citenamefont {Nakatsuji},\ and\ \citenamefont {Jungwirth}}]{Smejkal2022Jun}%
  \BibitemOpen
  \bibfield  {author} {\bibinfo {author} {\bibfnamefont {Libor}\ \bibnamefont {{\ifmmode\check{S}\else\v{S}\fi}mejkal}}, \bibinfo {author} {\bibfnamefont {Allan~H.}\ \bibnamefont {MacDonald}}, \bibinfo {author} {\bibfnamefont {Jairo}\ \bibnamefont {Sinova}}, \bibinfo {author} {\bibfnamefont {Satoru}\ \bibnamefont {Nakatsuji}}, \ and\ \bibinfo {author} {\bibfnamefont {Tomas}\ \bibnamefont {Jungwirth}},\ }\bibfield  {title} {\enquote {\bibinfo {title} {{Anomalous Hall antiferromagnets}},}\ }\href {\doibase 10.1038/s41578-022-00430-3} {\bibfield  {journal} {\bibinfo  {journal} {Nat. Rev. Mater.}\ }\textbf {\bibinfo {volume} {7}},\ \bibinfo {pages} {482--496} (\bibinfo {year} {2022}{\natexlab{c}})}\BibitemShut {NoStop}%
\bibitem [{\citenamefont {Ma}\ \emph {et~al.}(2021)\citenamefont {Ma}, \citenamefont {Hu}, \citenamefont {Li}, \citenamefont {Liu}, \citenamefont {Yao}, \citenamefont {Jia},\ and\ \citenamefont {Liu}}]{Ma2021May}%
  \BibitemOpen
  \bibfield  {author} {\bibinfo {author} {\bibfnamefont {Hai-Yang}\ \bibnamefont {Ma}}, \bibinfo {author} {\bibfnamefont {Mengli}\ \bibnamefont {Hu}}, \bibinfo {author} {\bibfnamefont {Nana}\ \bibnamefont {Li}}, \bibinfo {author} {\bibfnamefont {Jianpeng}\ \bibnamefont {Liu}}, \bibinfo {author} {\bibfnamefont {Wang}\ \bibnamefont {Yao}}, \bibinfo {author} {\bibfnamefont {Jin-Feng}\ \bibnamefont {Jia}}, \ and\ \bibinfo {author} {\bibfnamefont {Junwei}\ \bibnamefont {Liu}},\ }\bibfield  {title} {\enquote {\bibinfo {title} {{Multifunctional antiferromagnetic materials with giant piezomagnetism and noncollinear spin current}},}\ }\href {\doibase 10.1038/s41467-021-23127-7} {\bibfield  {journal} {\bibinfo  {journal} {Nat. Commun.}\ }\textbf {\bibinfo {volume} {12}},\ \bibinfo {pages} {1--8} (\bibinfo {year} {2021})}\BibitemShut {NoStop}%
\bibitem [{\citenamefont {McClarty}\ and\ \citenamefont {Rau}(2024)}]{McClarty2023Aug}%
  \BibitemOpen
  \bibfield  {author} {\bibinfo {author} {\bibfnamefont {Paul~A.}\ \bibnamefont {McClarty}}\ and\ \bibinfo {author} {\bibfnamefont {Jeffrey~G.}\ \bibnamefont {Rau}},\ }\bibfield  {title} {\enquote {\bibinfo {title} {{Landau Theory of Altermagnetism}},}\ }\href {\doibase 10.1103/PhysRevLett.132.176702} {\bibfield  {journal} {\bibinfo  {journal} {Phys. Rev. Lett.}\ }\textbf {\bibinfo {volume} {132}},\ \bibinfo {pages} {176702} (\bibinfo {year} {2024})}\BibitemShut {NoStop}%
\bibitem [{\citenamefont {Xiao}\ \emph {et~al.}(2024)\citenamefont {Xiao}, \citenamefont {Li}, \citenamefont {Han}, \citenamefont {Gan}, \citenamefont {Yang}, \citenamefont {Shao}, \citenamefont {Zhang}, \citenamefont {Gao}, \citenamefont {Tian},\ and\ \citenamefont {Zhou}}]{Xiao2024}%
  \BibitemOpen
  \bibfield  {author} {\bibinfo {author} {\bibfnamefont {Rui-Chun}\ \bibnamefont {Xiao}}, \bibinfo {author} {\bibfnamefont {Hui}\ \bibnamefont {Li}}, \bibinfo {author} {\bibfnamefont {Hui}\ \bibnamefont {Han}}, \bibinfo {author} {\bibfnamefont {Wei}\ \bibnamefont {Gan}}, \bibinfo {author} {\bibfnamefont {Mengmeng}\ \bibnamefont {Yang}}, \bibinfo {author} {\bibfnamefont {Ding-Fu}\ \bibnamefont {Shao}}, \bibinfo {author} {\bibfnamefont {Shu-Hui}\ \bibnamefont {Zhang}}, \bibinfo {author} {\bibfnamefont {Yang}\ \bibnamefont {Gao}}, \bibinfo {author} {\bibfnamefont {Mingliang}\ \bibnamefont {Tian}}, \ and\ \bibinfo {author} {\bibfnamefont {Jianhui}\ \bibnamefont {Zhou}},\ }\bibfield  {title} {\enquote {\bibinfo {title} {{Anomalous-Hall Neel textures in altermagnetic materials}},}\ }\href {\doibase 10.48550/arXiv.2411.10147} {\bibfield  {journal} {\bibinfo  {journal} {arXiv}\ } (\bibinfo {year} {2024}),\ 10.48550/arXiv.2411.10147},\ \Eprint {http://arxiv.org/abs/2411.10147} {2411.10147} \BibitemShut {NoStop}%
\bibitem [{\citenamefont {Fernandes}\ \emph {et~al.}(2024)\citenamefont {Fernandes}, \citenamefont {de~Carvalho}, \citenamefont {Birol},\ and\ \citenamefont {Pereira}}]{Fernandes2024Jan}%
  \BibitemOpen
  \bibfield  {author} {\bibinfo {author} {\bibfnamefont {Rafael~M.}\ \bibnamefont {Fernandes}}, \bibinfo {author} {\bibfnamefont {Vanuildo~S.}\ \bibnamefont {de~Carvalho}}, \bibinfo {author} {\bibfnamefont {Turan}\ \bibnamefont {Birol}}, \ and\ \bibinfo {author} {\bibfnamefont {Rodrigo~G.}\ \bibnamefont {Pereira}},\ }\bibfield  {title} {\enquote {\bibinfo {title} {{Topological transition from nodal to nodeless Zeeman splitting in altermagnets}},}\ }\href {\doibase 10.1103/PhysRevB.109.024404} {\bibfield  {journal} {\bibinfo  {journal} {Phys. Rev. B}\ }\textbf {\bibinfo {volume} {109}},\ \bibinfo {pages} {024404} (\bibinfo {year} {2024})}\BibitemShut {NoStop}%
\bibitem [{\citenamefont {Cheong}\ and\ \citenamefont {Huang}(2024)}]{Cheong2024Jan}%
  \BibitemOpen
  \bibfield  {author} {\bibinfo {author} {\bibfnamefont {Sang-Wook}\ \bibnamefont {Cheong}}\ and\ \bibinfo {author} {\bibfnamefont {Fei-Ting}\ \bibnamefont {Huang}},\ }\bibfield  {title} {\enquote {\bibinfo {title} {{Altermagnetism with non-collinear spins}},}\ }\href {\doibase 10.1038/s41535-024-00626-6} {\bibfield  {journal} {\bibinfo  {journal} {npj Quantum Mater.}\ }\textbf {\bibinfo {volume} {9}},\ \bibinfo {pages} {1--6} (\bibinfo {year} {2024})}\BibitemShut {NoStop}%
\bibitem [{\citenamefont {Mazin}\ and\ \citenamefont {Belashchenko}(2024)}]{Mazin2024Jul}%
  \BibitemOpen
  \bibfield  {author} {\bibinfo {author} {\bibfnamefont {I.~I.}\ \bibnamefont {Mazin}}\ and\ \bibinfo {author} {\bibfnamefont {K.~D.}\ \bibnamefont {Belashchenko}},\ }\bibfield  {title} {\enquote {\bibinfo {title} {Origin of the gossamer ferromagnetism in {MnTe}},}\ }\href {\doibase 10.1103/PhysRevB.110.214436} {\bibfield  {journal} {\bibinfo  {journal} {Phys. Rev. B}\ }\textbf {\bibinfo {volume} {110}},\ \bibinfo {pages} {214436} (\bibinfo {year} {2024})}\BibitemShut {NoStop}%
\bibitem [{\citenamefont {Berlijn}\ \emph {et~al.}(2017)\citenamefont {Berlijn}, \citenamefont {Snijders}, \citenamefont {Delaire}, \citenamefont {Zhou}, \citenamefont {Maier}, \citenamefont {Cao}, \citenamefont {Chi}, \citenamefont {Matsuda}, \citenamefont {Wang}, \citenamefont {Koehler}, \citenamefont {Kent},\ and\ \citenamefont {Weitering}}]{Berlijn2017Feb}%
  \BibitemOpen
  \bibfield  {author} {\bibinfo {author} {\bibfnamefont {T.}~\bibnamefont {Berlijn}}, \bibinfo {author} {\bibfnamefont {P.~C.}\ \bibnamefont {Snijders}}, \bibinfo {author} {\bibfnamefont {O.}~\bibnamefont {Delaire}}, \bibinfo {author} {\bibfnamefont {H.-D.}\ \bibnamefont {Zhou}}, \bibinfo {author} {\bibfnamefont {T.~A.}\ \bibnamefont {Maier}}, \bibinfo {author} {\bibfnamefont {H.-B.}\ \bibnamefont {Cao}}, \bibinfo {author} {\bibfnamefont {S.-X.}\ \bibnamefont {Chi}}, \bibinfo {author} {\bibfnamefont {M.}~\bibnamefont {Matsuda}}, \bibinfo {author} {\bibfnamefont {Y.}~\bibnamefont {Wang}}, \bibinfo {author} {\bibfnamefont {M.~R.}\ \bibnamefont {Koehler}}, \bibinfo {author} {\bibfnamefont {P.~R.~C.}\ \bibnamefont {Kent}}, \ and\ \bibinfo {author} {\bibfnamefont {H.~H.}\ \bibnamefont {Weitering}},\ }\bibfield  {title} {\enquote {\bibinfo {title} {{Itinerant Antiferromagnetism in ${\mathrm{RuO}}_{2}$}},}\ }\href {\doibase 10.1103/PhysRevLett.118.077201} {\bibfield  {journal} {\bibinfo  {journal} {Phys. Rev.
  Lett.}\ }\textbf {\bibinfo {volume} {118}},\ \bibinfo {pages} {077201} (\bibinfo {year} {2017})}\BibitemShut {NoStop}%
\bibitem [{\citenamefont {Hiraishi}\ \emph {et~al.}(2024)\citenamefont {Hiraishi}, \citenamefont {Okabe}, \citenamefont {Koda}, \citenamefont {Kadono}, \citenamefont {Muroi}, \citenamefont {Hirai},\ and\ \citenamefont {Hiroi}}]{Hiraishi2024Apr}%
  \BibitemOpen
  \bibfield  {author} {\bibinfo {author} {\bibfnamefont {M.}~\bibnamefont {Hiraishi}}, \bibinfo {author} {\bibfnamefont {H.}~\bibnamefont {Okabe}}, \bibinfo {author} {\bibfnamefont {A.}~\bibnamefont {Koda}}, \bibinfo {author} {\bibfnamefont {R.}~\bibnamefont {Kadono}}, \bibinfo {author} {\bibfnamefont {T.}~\bibnamefont {Muroi}}, \bibinfo {author} {\bibfnamefont {D.}~\bibnamefont {Hirai}}, \ and\ \bibinfo {author} {\bibfnamefont {Z.}~\bibnamefont {Hiroi}},\ }\bibfield  {title} {\enquote {\bibinfo {title} {{Nonmagnetic Ground State in ${\mathrm{RuO}}_{2}$ Revealed by Muon Spin Rotation}},}\ }\href {\doibase 10.1103/PhysRevLett.132.166702} {\bibfield  {journal} {\bibinfo  {journal} {Phys. Rev. Lett.}\ }\textbf {\bibinfo {volume} {132}},\ \bibinfo {pages} {166702} (\bibinfo {year} {2024})}\BibitemShut {NoStop}%
\bibitem [{\citenamefont {Keßler}\ \emph {et~al.}(2024)\citenamefont {Keßler}, \citenamefont {Garcia-Gassull}, \citenamefont {Suter}, \citenamefont {Prokscha}, \citenamefont {Salman}, \citenamefont {Khalyavin}, \citenamefont {Manuel}, \citenamefont {Orlandi}, \citenamefont {Mazin}, \citenamefont {Valentí},\ and\ \citenamefont {Moser}}]{Kessler2024May}%
  \BibitemOpen
  \bibfield  {author} {\bibinfo {author} {\bibfnamefont {Philipp}\ \bibnamefont {Keßler}}, \bibinfo {author} {\bibfnamefont {Laura}\ \bibnamefont {Garcia-Gassull}}, \bibinfo {author} {\bibfnamefont {Andreas}\ \bibnamefont {Suter}}, \bibinfo {author} {\bibfnamefont {Thomas}\ \bibnamefont {Prokscha}}, \bibinfo {author} {\bibfnamefont {Zaher}\ \bibnamefont {Salman}}, \bibinfo {author} {\bibfnamefont {Dmitry}\ \bibnamefont {Khalyavin}}, \bibinfo {author} {\bibfnamefont {Pascal}\ \bibnamefont {Manuel}}, \bibinfo {author} {\bibfnamefont {Fabio}\ \bibnamefont {Orlandi}}, \bibinfo {author} {\bibfnamefont {Igor~I.}\ \bibnamefont {Mazin}}, \bibinfo {author} {\bibfnamefont {Roser}\ \bibnamefont {Valentí}}, \ and\ \bibinfo {author} {\bibfnamefont {Simon}\ \bibnamefont {Moser}},\ }\bibfield  {title} {\enquote {\bibinfo {title} {Absence of magnetic order in {RuO}$_2$: insights from $\mu${SR} spectroscopy and neutron diffraction},}\ }\href {\doibase 10.1038/s44306-024-00055-y} {\bibfield  {journal} {\bibinfo  {journal}
  {npj Spintronics}\ }\textbf {\bibinfo {volume} {2}} (\bibinfo {year} {2024}),\ 10.1038/s44306-024-00055-y}\BibitemShut {NoStop}%
\bibitem [{\citenamefont {Autieri}\ \emph {et~al.}(2025)\citenamefont {Autieri}, \citenamefont {Sattigeri}, \citenamefont {Cuono},\ and\ \citenamefont {Fakhredine}}]{Autieri2023Dec}%
  \BibitemOpen
  \bibfield  {author} {\bibinfo {author} {\bibfnamefont {Carmine}\ \bibnamefont {Autieri}}, \bibinfo {author} {\bibfnamefont {Raghottam~M.}\ \bibnamefont {Sattigeri}}, \bibinfo {author} {\bibfnamefont {Giuseppe}\ \bibnamefont {Cuono}}, \ and\ \bibinfo {author} {\bibfnamefont {Amar}\ \bibnamefont {Fakhredine}},\ }\bibfield  {title} {\enquote {\bibinfo {title} {{Staggered Dzyaloshinskii-Moriya interaction inducing weak ferromagnetism in centrosymmetric altermagnets and weak ferrimagnetism in noncentrosymmetric altermagnets}},}\ }\href {\doibase 10.1103/PhysRevB.111.054442} {\bibfield  {journal} {\bibinfo  {journal} {Phys. Rev. B}\ }\textbf {\bibinfo {volume} {111}},\ \bibinfo {pages} {054442} (\bibinfo {year} {2025})}\BibitemShut {NoStop}%
\bibitem [{\citenamefont {Mazin}\ \emph {et~al.}(2021)\citenamefont {Mazin}, \citenamefont {Koepernik}, \citenamefont {Johannes}, \citenamefont {Gonz{\ifmmode\acute{a}\else\'{a}\fi}lez-Hern{\ifmmode\acute{a}\else\'{a}\fi}ndez},\ and\ \citenamefont {{\ifmmode\check{S}\else\v{S}\fi}mejkal}}]{Mazin2021Oct}%
  \BibitemOpen
  \bibfield  {author} {\bibinfo {author} {\bibfnamefont {Igor~I.}\ \bibnamefont {Mazin}}, \bibinfo {author} {\bibfnamefont {Klaus}\ \bibnamefont {Koepernik}}, \bibinfo {author} {\bibfnamefont {Michelle~D.}\ \bibnamefont {Johannes}}, \bibinfo {author} {\bibfnamefont {Rafael}\ \bibnamefont {Gonz{\ifmmode\acute{a}\else\'{a}\fi}lez-Hern{\ifmmode\acute{a}\else\'{a}\fi}ndez}}, \ and\ \bibinfo {author} {\bibfnamefont {Libor}\ \bibnamefont {{\ifmmode\check{S}\else\v{S}\fi}mejkal}},\ }\bibfield  {title} {\enquote {\bibinfo {title} {{Prediction of unconventional magnetism in doped FeSb$_2$}},}\ }\href {\doibase 10.1073/pnas.2108924118} {\bibfield  {journal} {\bibinfo  {journal} {Proc. Natl. Acad. Sci. U.S.A.}\ }\textbf {\bibinfo {volume} {118}},\ \bibinfo {pages} {e2108924118} (\bibinfo {year} {2021})}\BibitemShut {NoStop}%
\bibitem [{\citenamefont {Attias}\ \emph {et~al.}(2024)\citenamefont {Attias}, \citenamefont {Levchenko},\ and\ \citenamefont {Khodas}}]{Attias:2024}%
  \BibitemOpen
  \bibfield  {author} {\bibinfo {author} {\bibfnamefont {Lotan}\ \bibnamefont {Attias}}, \bibinfo {author} {\bibfnamefont {Alex}\ \bibnamefont {Levchenko}}, \ and\ \bibinfo {author} {\bibfnamefont {Maxim}\ \bibnamefont {Khodas}},\ }\bibfield  {title} {\enquote {\bibinfo {title} {Intrinsic anomalous hall effect in altermagnets},}\ }\href {\doibase 10.1103/PhysRevB.110.094425} {\bibfield  {journal} {\bibinfo  {journal} {Phys. Rev. B}\ }\textbf {\bibinfo {volume} {110}},\ \bibinfo {pages} {094425} (\bibinfo {year} {2024})}\BibitemShut {NoStop}%
\bibitem [{\citenamefont {Milivojevi{\ifmmode\acute{c}\else\'{c}\fi}}\ \emph {et~al.}(2024)\citenamefont {Milivojevi{\ifmmode\acute{c}\else\'{c}\fi}}, \citenamefont {Orozovi{\ifmmode\acute{c}\else\'{c}\fi}}, \citenamefont {Picozzi}, \citenamefont {Gmitra},\ and\ \citenamefont {Stavri{\ifmmode\acute{c}\else\'{c}\fi}}}]{Milivojevic2024May}%
  \BibitemOpen
  \bibfield  {author} {\bibinfo {author} {\bibfnamefont {Marko}\ \bibnamefont {Milivojevi{\ifmmode\acute{c}\else\'{c}\fi}}}, \bibinfo {author} {\bibfnamefont {Marko}\ \bibnamefont {Orozovi{\ifmmode\acute{c}\else\'{c}\fi}}}, \bibinfo {author} {\bibfnamefont {Silvia}\ \bibnamefont {Picozzi}}, \bibinfo {author} {\bibfnamefont {Martin}\ \bibnamefont {Gmitra}}, \ and\ \bibinfo {author} {\bibfnamefont {Sr{\dj}an}\ \bibnamefont {Stavri{\ifmmode\acute{c}\else\'{c}\fi}}},\ }\bibfield  {title} {\enquote {\bibinfo {title} {{Interplay of altermagnetism and weak ferromagnetism in two-dimensional RuF$_4$}},}\ }\href {\doibase 10.1088/2053-1583/ad4c73} {\bibfield  {journal} {\bibinfo  {journal} {2D Mater.}\ }\textbf {\bibinfo {volume} {11}},\ \bibinfo {pages} {035025} (\bibinfo {year} {2024})}\BibitemShut {NoStop}%
\bibitem [{\citenamefont {Roig}\ \emph {et~al.}(2024)\citenamefont {Roig}, \citenamefont {Kreisel}, \citenamefont {Yu}, \citenamefont {Andersen},\ and\ \citenamefont {Agterberg}}]{Roig2024Feb}%
  \BibitemOpen
  \bibfield  {author} {\bibinfo {author} {\bibfnamefont {Merc\`e}\ \bibnamefont {Roig}}, \bibinfo {author} {\bibfnamefont {Andreas}\ \bibnamefont {Kreisel}}, \bibinfo {author} {\bibfnamefont {Yue}\ \bibnamefont {Yu}}, \bibinfo {author} {\bibfnamefont {Brian~M.}\ \bibnamefont {Andersen}}, \ and\ \bibinfo {author} {\bibfnamefont {Daniel~F.}\ \bibnamefont {Agterberg}},\ }\bibfield  {title} {\enquote {\bibinfo {title} {{Minimal models for altermagnetism}},}\ }\href {\doibase 10.1103/PhysRevB.110.144412} {\bibfield  {journal} {\bibinfo  {journal} {Phys. Rev. B}\ }\textbf {\bibinfo {volume} {110}},\ \bibinfo {pages} {144412} (\bibinfo {year} {2024})}\BibitemShut {NoStop}%
\bibitem [{\citenamefont {Li}\ \emph {et~al.}(2024{\natexlab{b}})\citenamefont {Li}, \citenamefont {Zhang}, \citenamefont {Liu},\ and\ \citenamefont {Liu}}]{Li2024}%
  \BibitemOpen
  \bibfield  {author} {\bibinfo {author} {\bibfnamefont {Jiayu}\ \bibnamefont {Li}}, \bibinfo {author} {\bibfnamefont {Ao}~\bibnamefont {Zhang}}, \bibinfo {author} {\bibfnamefont {Yuntian}\ \bibnamefont {Liu}}, \ and\ \bibinfo {author} {\bibfnamefont {Qihang}\ \bibnamefont {Liu}},\ }\bibfield  {title} {\enquote {\bibinfo {title} {{Group Theory on Quasisymmetry and Protected Near Degeneracy}},}\ }\href {\doibase 10.1103/PhysRevLett.133.026402} {\bibfield  {journal} {\bibinfo  {journal} {Phys. Rev. Lett.}\ }\textbf {\bibinfo {volume} {133}},\ \bibinfo {pages} {026402} (\bibinfo {year} {2024}{\natexlab{b}})}\BibitemShut {NoStop}%
\bibitem [{\citenamefont {Guo}\ \emph {et~al.}(2022)\citenamefont {Guo}, \citenamefont {Hu}, \citenamefont {Putzke}, \citenamefont {Diaz}, \citenamefont {Huang}, \citenamefont {Manna}, \citenamefont {Fan}, \citenamefont {Shekhar}, \citenamefont {Sun}, \citenamefont {Felser}, \citenamefont {Liu}, \citenamefont {Bernevig},\ and\ \citenamefont {Moll}}]{Guo2022}%
  \BibitemOpen
  \bibfield  {author} {\bibinfo {author} {\bibfnamefont {Chunyu}\ \bibnamefont {Guo}}, \bibinfo {author} {\bibfnamefont {Lunhui}\ \bibnamefont {Hu}}, \bibinfo {author} {\bibfnamefont {Carsten}\ \bibnamefont {Putzke}}, \bibinfo {author} {\bibfnamefont {Jonas}\ \bibnamefont {Diaz}}, \bibinfo {author} {\bibfnamefont {Xiangwei}\ \bibnamefont {Huang}}, \bibinfo {author} {\bibfnamefont {Kaustuv}\ \bibnamefont {Manna}}, \bibinfo {author} {\bibfnamefont {Feng-Ren}\ \bibnamefont {Fan}}, \bibinfo {author} {\bibfnamefont {Chandra}\ \bibnamefont {Shekhar}}, \bibinfo {author} {\bibfnamefont {Yan}\ \bibnamefont {Sun}}, \bibinfo {author} {\bibfnamefont {Claudia}\ \bibnamefont {Felser}}, \bibinfo {author} {\bibfnamefont {Chaoxing}\ \bibnamefont {Liu}}, \bibinfo {author} {\bibfnamefont {B.~Andrei}\ \bibnamefont {Bernevig}}, \ and\ \bibinfo {author} {\bibfnamefont {Philip J.~W.}\ \bibnamefont {Moll}},\ }\bibfield  {title} {\enquote {\bibinfo {title} {{Quasi-symmetry-protected topology in a semi-metal}},}\ }\href {\doibase
  10.1038/s41567-022-01604-0} {\bibfield  {journal} {\bibinfo  {journal} {Nature Physics}\ }\textbf {\bibinfo {volume} {18}},\ \bibinfo {pages} {813--818} (\bibinfo {year} {2022})}\BibitemShut {NoStop}%
\bibitem [{\citenamefont {Hou}\ \emph {et~al.}(2023)\citenamefont {Hou}, \citenamefont {Yang}, \citenamefont {Liu}, \citenamefont {Guo},\ and\ \citenamefont {Lu}}]{Hou2023}%
  \BibitemOpen
  \bibfield  {author} {\bibinfo {author} {\bibfnamefont {Xiao-Yao}\ \bibnamefont {Hou}}, \bibinfo {author} {\bibfnamefont {Huan-Cheng}\ \bibnamefont {Yang}}, \bibinfo {author} {\bibfnamefont {Zheng-Xin}\ \bibnamefont {Liu}}, \bibinfo {author} {\bibfnamefont {Peng-Jie}\ \bibnamefont {Guo}}, \ and\ \bibinfo {author} {\bibfnamefont {Zhong-Yi}\ \bibnamefont {Lu}},\ }\bibfield  {title} {\enquote {\bibinfo {title} {{Large intrinsic anomalous Hall effect in both ${\mathrm{Nb}}_{2}{\mathrm{FeB}}_{2}$ and ${\mathrm{Ta}}_{2}{\mathrm{FeB}}_{2}$ with collinear antiferromagnetism}},}\ }\href {\doibase 10.1103/PhysRevB.107.L161109} {\bibfield  {journal} {\bibinfo  {journal} {Phys. Rev. B}\ }\textbf {\bibinfo {volume} {107}},\ \bibinfo {pages} {L161109} (\bibinfo {year} {2023})}\BibitemShut {NoStop}%
\bibitem [{\citenamefont {Roig}\ \emph {et~al.}()\citenamefont {Roig}, \citenamefont {Yu}, \citenamefont {Ekman}, \citenamefont {Kreisel}, \citenamefont {Andersen},\ and\ \citenamefont {Agterberg}}]{Supplementary}%
  \BibitemOpen
  \bibfield  {author} {\bibinfo {author} {\bibfnamefont {Mercè}\ \bibnamefont {Roig}}, \bibinfo {author} {\bibfnamefont {Yue}\ \bibnamefont {Yu}}, \bibinfo {author} {\bibfnamefont {Rune~C.}\ \bibnamefont {Ekman}}, \bibinfo {author} {\bibfnamefont {Andreas}\ \bibnamefont {Kreisel}}, \bibinfo {author} {\bibfnamefont {Brian~M.}\ \bibnamefont {Andersen}}, \ and\ \bibinfo {author} {\bibfnamefont {Daniel~F.}\ \bibnamefont {Agterberg}},\ }\href@noop {} {\enquote {\bibinfo {title} {{Supplementary Material for: Quasi-symmetry Constrained Spin Ferromagnetism in Altermagnets}},}\ }\BibitemShut {NoStop}%
\bibitem [{\citenamefont {Hecker}\ \emph {et~al.}(2024)\citenamefont {Hecker}, \citenamefont {Rastogi}, \citenamefont {Agterberg},\ and\ \citenamefont {Fernandes}}]{Hecker2024Jun}%
  \BibitemOpen
  \bibfield  {author} {\bibinfo {author} {\bibfnamefont {Matthias}\ \bibnamefont {Hecker}}, \bibinfo {author} {\bibfnamefont {Anant}\ \bibnamefont {Rastogi}}, \bibinfo {author} {\bibfnamefont {Daniel~F.}\ \bibnamefont {Agterberg}}, \ and\ \bibinfo {author} {\bibfnamefont {Rafael~M.}\ \bibnamefont {Fernandes}},\ }\bibfield  {title} {\enquote {\bibinfo {title} {{Classification of electronic nematicity in three-dimensional crystals and quasicrystals}},}\ }\href {\doibase 10.1103/PhysRevB.109.235148} {\bibfield  {journal} {\bibinfo  {journal} {Phys. Rev. B}\ }\textbf {\bibinfo {volume} {109}},\ \bibinfo {pages} {235148} (\bibinfo {year} {2024})}\BibitemShut {NoStop}%
\bibitem [{\citenamefont {Hayami}\ \emph {et~al.}(2018)\citenamefont {Hayami}, \citenamefont {Yatsushiro}, \citenamefont {Yanagi},\ and\ \citenamefont {Hiroaki}}]{Hayami2018Oct}%
  \BibitemOpen
  \bibfield  {author} {\bibinfo {author} {\bibfnamefont {Satoru}\ \bibnamefont {Hayami}}, \bibinfo {author} {\bibfnamefont {Megumi}\ \bibnamefont {Yatsushiro}}, \bibinfo {author} {\bibfnamefont {Yuki}\ \bibnamefont {Yanagi}}, \ and\ \bibinfo {author} {\bibfnamefont {Kusunose}\ \bibnamefont {Hiroaki}},\ }\bibfield  {title} {\enquote {\bibinfo {title} {{Classification of atomic-scale multipoles under crystallographic point groups and application to linear response tensors}},}\ }\href {\doibase 10.1103/PhysRevB.98.165110} {\bibfield  {journal} {\bibinfo  {journal} {Phys. Rev. B}\ }\textbf {\bibinfo {volume} {98}},\ \bibinfo {pages} {165110} (\bibinfo {year} {2018})}\BibitemShut {NoStop}%
\bibitem [{\citenamefont {Bhowal}\ and\ \citenamefont {Spaldin}(2024)}]{Bhowal2024Feb}%
  \BibitemOpen
  \bibfield  {author} {\bibinfo {author} {\bibfnamefont {Sayantika}\ \bibnamefont {Bhowal}}\ and\ \bibinfo {author} {\bibfnamefont {Nicola~A.}\ \bibnamefont {Spaldin}},\ }\bibfield  {title} {\enquote {\bibinfo {title} {{Ferroically Ordered Magnetic Octupoles in $d$-Wave Altermagnets}},}\ }\href {\doibase 10.1103/PhysRevX.14.011019} {\bibfield  {journal} {\bibinfo  {journal} {Phys. Rev. X}\ }\textbf {\bibinfo {volume} {14}},\ \bibinfo {pages} {011019} (\bibinfo {year} {2024})}\BibitemShut {NoStop}%
\bibitem [{\citenamefont {Graf}\ and\ \citenamefont {Pi{\ifmmode\acute{e}\else\'{e}\fi}chon}(2021)}]{Graf2021Aug}%
  \BibitemOpen
  \bibfield  {author} {\bibinfo {author} {\bibfnamefont {Ansgar}\ \bibnamefont {Graf}}\ and\ \bibinfo {author} {\bibfnamefont {Fr{\ifmmode\acute{e}\else\'{e}\fi}d{\ifmmode\acute{e}\else\'{e}\fi}ric}\ \bibnamefont {Pi{\ifmmode\acute{e}\else\'{e}\fi}chon}},\ }\bibfield  {title} {\enquote {\bibinfo {title} {{Berry curvature and quantum metric in $N$-band systems: An eigenprojector approach}},}\ }\href {\doibase 10.1103/PhysRevB.104.085114} {\bibfield  {journal} {\bibinfo  {journal} {Phys. Rev. B}\ }\textbf {\bibinfo {volume} {104}},\ \bibinfo {pages} {085114} (\bibinfo {year} {2021})}\BibitemShut {NoStop}%
\bibitem [{\citenamefont {Jo}\ \emph {et~al.}(2024)\citenamefont {Jo}, \citenamefont {Go}, \citenamefont {Mokrousov}, \citenamefont {Oppeneer}, \citenamefont {Cheong},\ and\ \citenamefont {Lee}}]{J02024}%
  \BibitemOpen
  \bibfield  {author} {\bibinfo {author} {\bibfnamefont {Daegeun}\ \bibnamefont {Jo}}, \bibinfo {author} {\bibfnamefont {Dongwook}\ \bibnamefont {Go}}, \bibinfo {author} {\bibfnamefont {Yuriy}\ \bibnamefont {Mokrousov}}, \bibinfo {author} {\bibfnamefont {Peter~M.}\ \bibnamefont {Oppeneer}}, \bibinfo {author} {\bibfnamefont {Sang-Wook}\ \bibnamefont {Cheong}}, \ and\ \bibinfo {author} {\bibfnamefont {Hyun-Woo}\ \bibnamefont {Lee}},\ }\bibfield  {title} {\enquote {\bibinfo {title} {{Weak Ferromagnetism in Altermagnets from Alternating $g$-Tensor Anisotropy}},}\ }\href {\doibase 10.48550/arXiv.2410.17386} {\bibfield  {journal} {\bibinfo  {journal} {arXiv}\ } (\bibinfo {year} {2024}),\ 10.48550/arXiv.2410.17386},\ \Eprint {http://arxiv.org/abs/(accepted in PRL)} {(accepted in PRL)} \BibitemShut {NoStop}%
\end{thebibliography}%

\begin{figure}[tb]
\begin{center}
\includegraphics[angle=0,width=.95\linewidth]{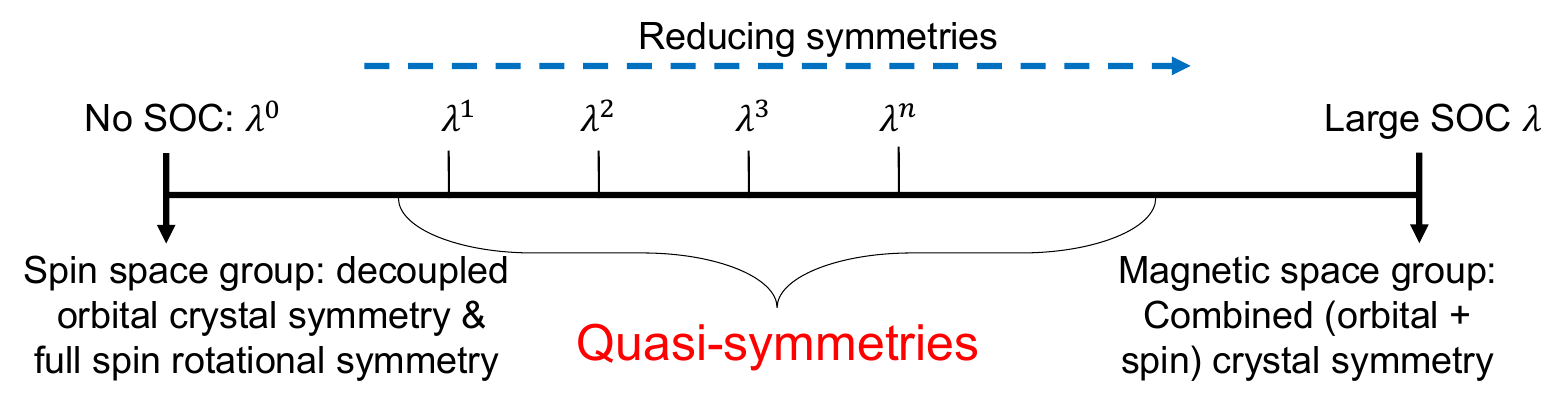}
\caption{Classification scheme of the spin space group and magnetic space group in terms of the strength and powers of SOC, illustrating when the quasi-symmetries emerge. In particular, when two of the SOC components vanish, the resulting symmetry group (which we denote as the uniaxial spin space group) has spin rotational invariance around the SOC direction.}
\label{fig:quasisym_illustration}
\end{center}
\end{figure}

\appendix
\begin{widetext}
\section{End Matter}
\end{widetext}
\textit{{\red Appendix: General SOC-enabled quasi-symmetry--}} As introduced in the main text, the quasi-symmetry generated by SOC when two of the SOC components vanish allows us to determine which Landau coefficients are linear in one of $\lambda_{x,y,z}$. As illustrated in Fig.~\ref{fig:quasisym_illustration}, the resulting symmetry group, denoted here as the uniaxial spin space group, has lower symmetry than the spin space group, but higher symmetry than the magnetic space group.
Here, we develop a more general quasi-symmetry criterion that addresses the following question: can a Landau coefficient or response function host a contribution from $\lambda_x^{n_x} \lambda_y^{n_y} \lambda_z^{n_z}$, for given integers $(n_x,n_y,n_z)\geq0$?


We would like to consider a general Hamiltonian that can extend beyond the minimal models discussed in this paper. SOC terms can always be grouped into terms proportional to spin operators $\sigma_x$, $\sigma_y$, and $\sigma_z$. Terms in each group can carry different momentum and orbital dependence, with overall SOC strengths $\lambda_x$, $\lambda_y$, and $\lambda_z$:
\begin{align}
H_{SOC}=& \, \ \lambda_x(a_{1,{\bf k}}\hat{A_1}+a_{2,{\bf k}}\hat{A_2}+...)\sigma_x \nonumber \\
&+\lambda_y(b_{1,{\bf k}}\hat{B_1}+b_{2,{\bf k}}\hat{B_2}+...)\sigma_y \nonumber \\
&+\lambda_z(c_{1,{\bf k}}\hat{C_1}+c_{2,{\bf k}}\hat{C_2}+...)\sigma_z,
\end{align}
with momentum-dependent coefficients $a_i,b_i,c_i=O(1)$ and $\hat{A_i}$, $\hat{B_i}$, $\hat{C_i}$ some operators acting on sublattice and/or orbital space.
The quasi-symmetry argument is based on analyticity of the response functions. If the DOS is not concentrated at singular regions such as band crossings, response functions should be an analytical function of $\lambda_{x,y,z}$. Cross-terms like $\sqrt{\lambda_x\lambda_y}$ are not allowed. For the minimal model, this property is explicitly shown in the normal-state Green's function Eq.~\eqref{eq:G_projected_bandbasis}, where the projection operator is a linear combination of the three SOC terms.

We now interpret the three groups of SOC terms as distinct symmetry-breaking order parameters in a SOC-free system (spin-space group). The spin-space group hosts decoupled orbital crystal symmetries and full spin-rotational symmetries. General order parameters are denoted by $\vec{O}_{R,i}$, where R is the IR under orbital rotations and the index $i = 1, 2, \dots$ reflects its dimensionality. The vector notation indicates that each $\vec{O}_{R,i}$ also transforms as a spin vector under spin rotations.

Each group of SOC terms corresponds to a different symmetry breaking in this high-symmetry system. For example, consider the $\lambda_z$ SOC term in an tetragonal system. It breaks spin-rotational symmetry as the spin vector $\sigma_z$, and it breaks orbital symmetry according to the 1D IR $A_{2g}\sim xy(x^2-y^2)$. It is then $\vec{O}_{A_{2g}}=(0,0,\Lambda_z)$. For $\lambda_{x,y}$ SOC terms, they break spin-rotational symmetry as the spin vector $\sigma_{x,y}$, and they break orbital symmetry according to the 2D IR $E_{g}\sim \{xz,yz\}$. They are then $\vec{O}_{E_{g},1}=(\Lambda_{xy},0,0)$ and $\vec{O}_{E_{g},2}=(0,\Lambda_{xy},0)$.

If the coefficient of a Landau term or response function $X$ hosts a contribution from $\lambda_x^{n_x}\lambda_y^{n_y}\lambda_z^{n_z}$, then the term $\Lambda_x^{n_x}\Lambda_y^{n_y}\Lambda_z^{n_z}X$ must be allowed in the Landau theory for the SOC-free system.
We note that for the tetragonal system, the contributions from $\lambda_{x,y}^{n_{xy}}\lambda_z^{n_z}$ enable the term  $\Lambda_{xy}^{n_{xy}}\Lambda_z^{n_z}X$ in the Landau theory for the SOC-free system (since $\lambda_x = \lambda_y$ in this case).
In the SOC-free system, a valid Landau term should be a spin-scalar, and belong to the trivial IR under orbital rotations. This provides a general quasi-symmetry criterion.

\textit{{\red Examples--}} 
\begin{enumerate}
    \item In the SOC-free tetragonal $D_{4h}$ systems, the SOC order parameters are: $\vec{O}_{E_{g},1}=(\Lambda_{xy},0,0)$, $\vec{O}_{E_{g},2}=(0,\Lambda_{xy},0)$ and  $\vec{O}_{A_{2g}}=(0,0,\Lambda_z)$. Ferromagnetic order is $\vec{M}_{A_{1g}}$, and altermagnetic order is $\vec{N}_{P}$, with altermagnetic symmetry $\Gamma_N = P$. To have a SOC-linear coupling between an altermagnet and a ferromagnet, $\vec{O}\cdot(\vec{M}\times\vec{N})$ must be allowed. The altermagnetic symmetry thus has to be $A_{2g}$ or $E_g$. For two atoms per nonmagnetic unit cell, only $\Gamma_N = A_{2g}, B_{1g}, B_{2g}$ are allowed~\cite{Roig2024Feb}.
    \begin{enumerate}
        \item For $\Gamma_N = A_{2g}$, the Landau term $\vec{O}_{A_{2g}}\cdot(\vec{M}_{A_{1g}}\times\vec{N}_{A_{2g}})=\Lambda_z(M_xN_y-M_yN_x)$ is allowed. Hence the $\lambda_z$-linear contribution to $M_xN_y-M_yN_x$ is nonzero. 
        \item For $\Gamma_N = B_{2g}$ and Landau term $M_xN_y+M_yN_x$, SOC-linear terms are forbidden. Since $E_g\otimes E_g\otimes B_{2g}$ contains the trivial IR, the quadratic $\lambda_{xy}^2$ contribution is allowed. An example Landau term is $(\vec{O}_{E_{g},1}\times\vec{M})\cdot(\vec{N}\times\vec{O}_{E_{g},2})+(\vec{O}_{E_{g},2}\times\vec{M})\cdot(\vec{N}\times\vec{O}_{E_{g},1})=\Lambda_{xy}^2(M_xN_y+M_yN_x)$. 
        \item For $\Gamma_N = B_{1g}$ and Landau term $M_xN_x-M_yN_y$, SOC-linear terms are forbidden. Since $E_g\otimes E_g\otimes B_{1g}$ contains the trivial IR, the quadratic $\lambda_{xy}^2$ contribution is allowed. An example Landau term is $(\vec{O}_{E_{g},1}\times\vec{M})\cdot(\vec{N}\times\vec{O}_{E_{g},1})-(\vec{O}_{E_{g},2}\times\vec{M})\cdot(\vec{N}\times\vec{O}_{E_{g},2})=\Lambda_{xy}^2(M_xN_x-M_yN_y)$. 
    \end{enumerate}
    \item In the SOC-free hexagonal $D_{6h}$ systems, the SOC order parameters are: $\vec{O}_{E_{1g},1}=(\Lambda_{xy},0,0)$, $\vec{O}_{E_{1g},2}=(0,\Lambda_{xy},0)$ and  $\vec{O}_{A_{2g}}=(0,0,\Lambda_z)$. To have a SOC-linear coupling between altermagnet and ferromagnet, $\vec{O}\cdot(\vec{M}\times\vec{N})$ must be allowed. Thus, $\Gamma_N = A_{2g}$ or $E_{1g}$. For two atoms per nonmagnetic unit cell, only $A_{2g}$, $B_{1g}$, and $B_{2g}$ are allowed~\cite{Roig2024Feb}.
    \begin{enumerate}
        \item The discussion for $\Gamma_N = A_{2g}$ is the same as $A_{2g}$ in tetragonal systems. 
        \item For $\Gamma_N = B_{1g}$ (similar for $\Gamma_N = B_{2g}$), $MN^3$ coupling has no SOC-linear contribution since $B_{1g}\otimes B_{1g}\otimes B_{1g}\otimes\{E_{1g},A_{2g}\}$ has no trivial IR. Similarly, SOC-quadratic contribution is also forbidden since $B_{1g}\otimes B_{1g}\otimes B_{1g}\otimes \{E_{1g},A_{2g}\}\otimes \{E_{1g},A_{2g}\}$ has no trivial IR. SOC-cubic contribution is allowed as $B_{1g}\otimes B_{1g}\otimes B_{1g}\otimes E_{1g}\otimes E_{1g}\otimes E_{1g}$ has trivial IR. The SOC dependence is cubic in $\lambda_{xy}$.
    \end{enumerate}
    \item In the SOC-free cubic $O_h$ systems, the SOC order parameters are: $\vec{O}_{T_{1g},1}=(\Lambda,0,0)$, $\vec{O}_{T_{1g},2}=(0,\Lambda,0)$ and  $\vec{O}_{T_{1g},3}=(0,0,\Lambda)$. For $\Gamma_N = A_{2g}$, $MN^3$ coupling has no SOC-linear contribution since $A_{2g}\otimes A_{2g}\otimes A_{2g}\otimes T_{1g}$ has no trivial IR. Similarly, a SOC-quadratic contribution is also forbidden since $A_{2g}\otimes A_{2g}\otimes A_{2g}\otimes T_{1g}\otimes T_{1g}$ has no trivial IR. Hence, the leading contribution is cubic in SOC, as $A_{2g}\otimes A_{2g}\otimes A_{2g}\otimes T_{1g}\otimes T_{1g}\otimes T_{1g}$ contains the trivial IR. 
\end{enumerate}

\ifarXiv
    \foreach \x in {1,...,\numbersupplementpages}
    {
        \clearpage
        \includepdf[pages={\x,{}}]{\supplementfilename}
    }
\fi
\end{document}